\documentclass[useAMS,usenatbib]{mn2e}

\usepackage{amsmath}
\usepackage{mathtext,amssymb,amsmath}
\usepackage{epsfig}
\usepackage{graphics}
\usepackage{url}
\usepackage{times}
\usepackage[T1]{fontenc} 
\usepackage{aecompl}

\renewcommand{\div}{\ensuremath{-}}

\newcommand{\bea}{\begin{eqnarray}}
\newcommand{\eea}{\end{eqnarray}}

\newcommand{\ergl}{\ensuremath{\,\rm erg\, s^{-1}}}

\newcommand{\mum}{\ensuremath{\,\rm \mu m}}
\newcommand{\pc}{\ensuremath{\,\rm pc}}
\newcommand{\cmc}{\ensuremath{\,\rm cm^{-3}}}
\newcommand{\cmsq}{\ensuremath{\,\rm cm^{-2}}}

\newcommand{\Ry}{\ensuremath{\,{\rm Ry}}}
\newcommand{\eV}{\ensuremath{\,{\rm eV}}}
\newcommand{\keV}{\ensuremath{\,{\rm keV}}}
\newcommand{\GeV}{\ensuremath{\,{\rm GeV}}}
\newcommand{\TeV}{\ensuremath{\,{\rm TeV}}}
\newcommand{\AAA}{\ensuremath{\,{\rm \AA}}}

\newcommand\ion[2]{#1$\,${\scshape{#2}}}

\newcommand{\cloudy}{{\sc cloudy}}
\newcommand{\xstar}{{\sc xstar}}

\title[Gamma-ray opacity of the BLR]{Gamma-ray opacity of the anisotropic stratified broad-line regions in
blazars}
\author[]{Pavel Abolmasov$^{1,2}$\thanks{E-mail:
pavel.abolmasov@gmail.com} and Juri Poutanen$^{1,3}$ \\
$^{1}$Tuorla Observatory, Department of Physics and Astronomy, University of Turku, V\"ais\"al\"antie 20, FI-21500
Piikki\"o, Finland,\\
$^{2}$Sternberg Astronomical Institute, Moscow State University,
  Universitetsky pr., 13, Moscow 119992, Russia\\
$^{3}$Nordita, KTH Royal Institute of Technology and Stockholm University, Roslagstullsbacken 23, SE-10691 Stockholm, Sweden
}

\begin{document}

\date{Accepted ---. Received ---; in
  original form --- }

\label{firstpage}
\pagerange{\pageref{firstpage}--\pageref{lastpage}} \pubyear{2016}
\maketitle

\begin{abstract}
The GeV-range spectra of blazars are shaped not only by non-thermal emission processes internal to the relativistic jet but also by external pair-production absorption on the thermal emission of the accretion disc and the broad-line region (BLR). 
For the first time, we compute here the pair-production opacities in the GeV range produced by a realistic BLR accounting for the radial stratification and radiation anisotropy. 
Using photoionization modelling with the \cloudy\ code, we calculate a series of BLR models of different sizes,  geometries, cloud densities, column densities and metallicities. 
The strongest emission features in the model BLR are Ly$\alpha$ and
\ion{He}{II}\,Ly$\alpha$. 
Contribution of recombination continua is smaller, especially for hydrogen,
because Ly continuum is efficiently trapped inside the large optical depth BLR
clouds and converted to Lyman emission lines and higher-order recombination
continua. 
The largest effects  on the gamma-ray opacity are produced by the BLR geometry
and localization of the gamma-ray source. 
We show that when the gamma-ray source moves further from the central source,
all the absorption details move to higher energies and the overall level of
absorption drops because of decreasing incidence angles between the gamma-rays and BLR
photons. 
The observed positions of the spectral breaks can be used to
measure the geometry and the location of the gamma-ray emitting region
relative to the BLR. 
Strong dependence on geometry means that the soft photons dominating the
pair-production opacity may be actually produced by a different population of
BLR clouds than the bulk of the observed broad line emission.
\end{abstract}


\section{Introduction}

Some gamma-ray source may be also bright at softer energies, so that  this soft radiation  becomes a source of opacity for the gamma-rays 
through photon-photon electron-positron pair production. 
In particular, radiation from the infrared to the EUV band (0.1--100~eV) contributes to the 
opacity in the 1--10$^3$~GeV range. 
This may be important for accreting black holes with gamma-ray emitting jets, both in close binary systems and in active galactic nuclei (AGN) as well as for pulsars with high-mass companions. 
Here we consider in detail the case of flat spectrum radio quasars (FSRQ), which  are not only  bright non-thermal sources from radio to gamma-ray energies 
but also powerful emitters of thermal optical/UV/EUV radiation. 
In this spectral range, isotropic (unbeamed) emission of a bright AGN is dominated by the so-called big blue bump (BBB) with a maximum around 1000\AA. 
Emission of the BBB is continual, but considerable part of the radiation comes in broad components of emission lines. Unlike the broad-band emission, the broad emission lines and recombination edges may produce relatively sharp spectral details in gamma-ray absorption. In particular, the strong Ly$\alpha$ line should create a spectral break at the threshold energy of about 25~GeV. 

Observational data hint that such absorption details do indeed exist. 
The spectra of FSRQ and bright (low-synchrotron-peak) BL\,Lacs obtained by the Large Area Telescope onboard of the {\it Fermi  Gamma-ray Space Telescope} ({\it Fermi}/LAT) in the 0.1\div30~GeV energy range reveal strong deviations from any smooth (single power-law or log-normal) spectral model \citep{fermi10, PS10,tanaka11,SP11,SP14}. 
Unlike the fainter BL\,Lac objects lacking observable disc and line emission, a nearly power-law spectrum observed in FSRQs in the $\lesssim 1$~GeV range cannot be extrapolated to energies higher than several GeV. 
The spectral slope  becomes steeper near a break energy of about several GeV. 
Qualitatively such details are well explained by pair-production opacity created by individual bright spectral lines and sharp spectral edges \citep{PS10,SP14} in the far and extreme UV range. 
Such spectral details are naturally produced by the BLR responsible for the broad components of emission lines. 
The best candidates for absorption at several GeV are the \ion{He}{ii}~Ly$\alpha$ line and Lyman recombination continuum (LyC). 
Hydrogen Ly$\alpha$ and LyC emission should contribute at $\sim 20\div 30\GeV$. 
However, because of the lack of photons above 20 GeV, it is challenging to judge about spectral shape at these energies.
The positions and even the existence of the breaks at several GeV have been questioned  \citep{Harris}. 
Some of the originally detected sharp features are likely the  artefacts of the {\it Fermi}/LAT Pass 6 response function, but still  some breaks are significantly detected in the redshift-corrected stacked spectra of blazars as well as in the spectra of individual bright sources in the Pass 7 data \citep{SP14}. 

The BLR is composed of dense ($n_{\rm H}\sim 10^{9}\div 10^{13}\cmc$) clumpy photoionized gas moving nearly chaotically at random velocities close to virial. 
The physical conditions in BLR are constrained by relative line intensities (see \citet{osterbrock}, sections 13.6 and 14.5, and references therein).
BLR gas is often viewed as some sort of a wind produced by the accretion disc (see \citealt{botorff97} and references therein) though there is strong observational evidence for inward-directed motions in BLRs \citep{doro12, grier13}. 
The BLR size measured through reverberation mapping depends on the quasar
UV luminosity $L_{\rm UV}$ as $R_{\rm  BLR,17}\sim 1\times
L_{\rm UV, 45}^{1/2}$\citep{kaspi07}.\footnote{Here and below we use notation
  $Q=10^x Q_x$ in cgs units. }
To reproduce the observed BLR spectra, individual clouds should be optically thick to Lyman continuum and the Lyman series lines. 
This means that the emission of BLR clumps should be highly anisotropic, and the inward- and outward-directed diffuse continua should differ considerably  \citep{anisofer}. 

The typical density of BLR photons within the BLR is  $n_{\rm ph,BLR}\approx 10^9\cmc$ independently of the luminosity. 
Here we assumed that 10 per cent of quasar luminosity is reprocessed in the BLR to photons of average energy 10 eV. 
The photon density is uncertain by  an order of magnitude, because of the uncertainties on the BLR radius determined using different emission lines \citep{PW99,MS12}. 

The gamma-ray radiation is believed to be produced by the relativistic jet and, therefore, is highly beamed in the direction of the jet propagation. 
In  quasars, this radiation would propagate through the radiation field of the accretion disc,  BLR and the dusty torus. 
While the photon density around quasar is dominated by the accretion disc, it is not the dominant source of gamma-ray opacity, as this radiation streams along the jet basically in the same direction as the gamma-rays leading to a strong reduction of the interaction rate. 
On the other hand, the BLR photons, distributed more or less isotropically,  collide with the beamed gamma-rays at much larger angles and, therefore, provide a much high opacity. 
For a gamma-ray source well inside a  BLR and photons above 30~GeV, we can
estimate the maximal optical depth created by the BLR emission (mostly by Ly
$\alpha$) as  $\tau_{\gamma\gamma} \sim 0.2 \sigma_{\rm T} n_{\rm ph,BLR}
R_{\rm BLR} \sim (1\div 100) L_{\rm UV, 45}^{1/2}$.

The first attempts to compute the gamma-ray opacity were made by \citet{liu_bai06} and \citet{Reimer07}, who considered sources of all observed broad lines distributed inside a spherical BLR shell. 
It is known, however, that lines of different ionization are produced at very different distances from the  central source \citep{PW99} and the gamma-ray opacity strongly depends on the local BLR UV spectrum which  cannot be directly  observed because of the internal as well as external absorption. 
This justifies the attempts to use photoionization models to predict the local BLR spectrum and, consequently, the gamma-ray absorption.  
Such modelling was made by \citet{tavecchio08} using \cloudy\ (the latest release of this photoionization code is described in \citealt{cloudy13}) and by \citet{PS10} using \xstar\ (see \citealt{xstar}). 
\citet{tavecchio12} have noticed a strong dependence of the gamma-ray opacity on the BLR geometry, within a framework of a single thin spherical shell (or its fraction).   
Similar conclusions were reached by \citet{LeiWang14PASJ} who considered the sources of the observed line emission.   
 
In this paper, we compute for the first time the photon-photon pair-production absorption through a realistic BLR. 
We consider an axisymmetric model for the anisotropically emitting, radially-stratified  BLR of different geometries.
In Section~\ref{sec:model}, the \cloudy\ BLR model is introduced. 
In Section~\ref{sec:tau}, we describe the calculation of the gamma-ray absorption. 
Results are given in Section~\ref{sec:res} and discussed in Section~\ref{sec:disc}. 

\section{Model of the BLR}\label{sec:model}

\subsection{Model setup}\label{sec:model:setup}

\begin{figure}
 \centering
\includegraphics[width=\columnwidth]{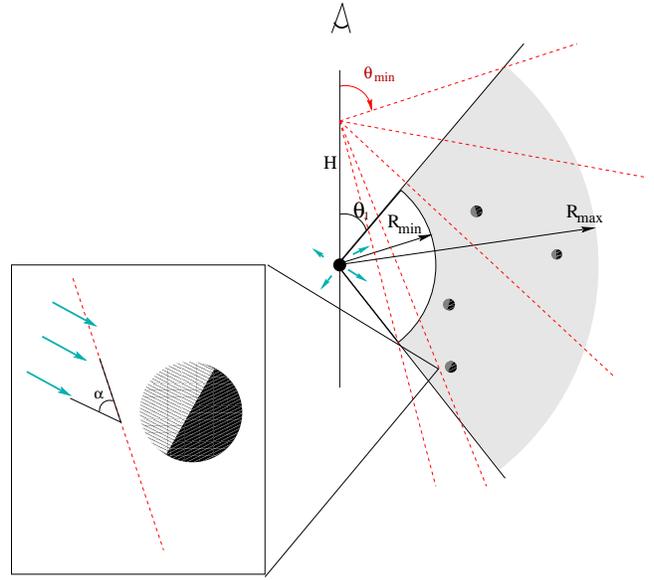}
\caption{A sketch illustrating the structure of BLR. The BLR is confined between the two spherical surfaces with radii $R_{\min}$ and $R_{\max} $ and between the conical surfaces of $\theta = \theta_1$ and $\theta=\pi-\theta_1$. 
The gamma-ray source is located above the centre at the altitude $H$. 
Dashed red lines are sample lines of integration, that intersect at the point where the radiation field is calculated.
}
\label{fig:scheme}
\end{figure}

Let us consider small and dense clouds distributed uniformly between some inner radius of $R_{\rm in}$ and outer radius $R_{\rm out}$ and in a range of polar angles from $\theta_1$ to $\pi-\theta_1$ (see Fig.~\ref{fig:scheme}), where $\theta_1$ may vary between $0$ (spherical symmetry) and $\pi/2$ (flat disc).
The region itself is considered optically thin in a sense that, for a fixed line of sight,  individual clouds practically never overlap. 
This allows us to use a single ionizing continuum shape for clouds at different distances from the source located in the centre and identified with the accretion disc. 
Ionizing continuum shape was taken in the conventional form (see
\citet{moloney} and references therein)
\begin{equation}\label{E:isp}
L_E \propto \left\{
\begin{array}{lc}
E^{5/2}, & \mbox{for $E<0.12$\eV,}\\
E^{-1}, & \mbox{for $0.12$\eV$<E<4$\Ry,}\\
E^{-2}, & \mbox{for $4$\Ry$<E<30$\Ry,}\\
E^{-1}, & \mbox{for $30$\Ry$<E<100$\keV,}\\
E^{-2}, & \mbox{for $E>100$\keV.}\\
\end{array}
\right.
\end{equation}
Such spectrum corresponds to the production rate of hydrogen-ionizing photons of $Q_{\rm ion}\simeq 2.6\times 10^{54} L_{45}$~s$^{-1}$. 
The model spectrum does not necessarily coincide with the observed thermal emission spectrum dominated by the BBB as we perceive. 
The observed BBB emission  may be some sort of reprocessed emission (see \citealt{lawrence12}), and is irrelevant as a source of ionizing photons because of its softness.
Position of the low-energy break was set according to \citet{anisofer} who found it to be influential upon the equilibrium temperature of the gas. 
The  position of the break around $100\mum$ reproduces the characteristic line
ratios like \ion{C}{IV}$\lambda$1549/Ly$\alpha$ and
\ion{O}{VI}$\lambda$1034/Ly$\alpha$. 

The source of the non-thermal emission may be comparable in power but beamed away from the BLR hence we will rather identify the source with the accretion disc.
Total luminosity was set to the moderate value of $L_{\rm tot} = 10^{45}\ergl$. 
For a large selection of FSRQ, \citet{delia03} estimate the luminosities of the accretion discs in the range from $10^{44}$ to several times $10^{47}\ergl$. 
In our assumptions, the simple scalings for the luminosity, radius and optical depth (see Section~\ref{sec:scale}) allow to apply the results of our calculations to different values of disc luminosity. 
We assume that the density $n_{\rm H}$ is constant inside a cloud and is identical for all the clouds. 
As we will see in Section~\ref{sec:blr}, this is probably the strongest over-simplification of our model, as co-existence of different emission lines like \ion{Mg}{II}$\lambda$2800 and \ion{He}{II}$\lambda$1640 requires not  only considerable spread in radii but also a span in densities. 
The commonly accepted value for hydrogen column density of individual clouds of $N_{\rm H} = 10^{23}\cmsq$  represents the characteristic or minimal hydrogen column density. 
Higher column densities provide slightly better fits to the observed line ratios due to existence of an extended region of partial ionization, producing lower-ionization emission lines. 

For each set of global parameters, $n_{\rm H}$ and $N_{\rm H}$, we calculate a grid of \cloudy\ models  for distances from the source between $10^{-3}$ to 1\pc. 
The BLR spectrum strongly depends on the ionization parameter 
\begin{equation}
U=\dfrac{Q_{\rm ion}}{4\pi c R^2n_{\rm H}} .
\end{equation}
The considered distances correspond to the ionization parameter in the range from $10^{-5}$ up to $10^2$.
This range covers well the ionization parameters $10^{-3}\div 10^{-1}$ estimated by ionization modelling of the observed BLR UV spectra  \citep{ruff12,negrete13} as well as the higher ionization parameters $\sim 1\div 100$  estimated from fitting of the gamma-ray absorption details \citep{PS10,SP14}. 
Large ionization parameters make ionization degree in the model cloud very high, and the spectrum of the illuminated side of the cloud becomes dominated by Thomson reflection. 

An example of an individual \cloudy\ input is given in Appendix~\ref{sec:app:cloudy}. 
Each cloud was treated as a one-dimensional constant density slab having hydrogen column density equal to some constant quantity $N_{\rm H}$, normally $N_{\rm H} = 10^{23}\cmsq$. 
Its transverse size was assumed equal to its radial size, $N_{\rm H} / n_{\rm H}$.
Different estimates exist for the density of BLR material, mostly in the range $n_{\rm H}=10^{9}\div 10^{13}\cmc$. 
The lower density limit of $n_{\rm H} \gtrsim 10^8\cmc$ is due to absence of forbidden lines in the spectrum. 
The existence and the brightness of the semi-forbidden emission \ion{C}{III]}$\lambda$1909 allows to estimate the density as $\sim  10^{10}\cmc$ \citep{osterbrock}. 
Higher density values are favoured by \citet{moloney}, \citet{negrete14} and \citet{ruff12} who fitted the existing hydrogen line   intensities with \cloudy\ models to reconstruct the spectral shape of the source. 
They also arrive to a rather large value of ionization parameter $U\sim 0.3$. 
Note that gas denser than $\sim 10^{13}\cmc$ does not contribute to intercombination lines like \ion{C}{III]}$\lambda$1909. 
Besides, it lies close to the applicability limits of the approximations used by \cloudy\ \citep{cloudy13}. 
As we will see below, emission produced by a thick slab with $N_{\rm H} \sim 10^{23}\cmsq$ is strongly anisotropic, hence it is important to take into account the angle at which the cloud is seen.  
BLR clouds are expected to emit strongly anisotropically \citep{anisofer}, therefore we consider the outward- and the inward-emitted radiation separately. 
Emission of an individual cloud can be approximated by emission of a sphere one half of which radiates the outward, the other the inward-directed (reflected) flux from the \cloudy\ output. 
The observed luminosity produced by the cloud in a unit solid angle is
\begin{equation}
\dfrac{dL_1}{d\Omega} = \displaystyle \pi R_c^2 \left( \frac{1+\cos \alpha}{2} F_{\rm in}(R) +  \frac{1-\cos \alpha}{2} F_{\rm out}(R)\right),
\end{equation}
where $R_c= N_{\rm H} / 2n_{\rm H} $ is the radius of the cloud, $F_{\rm in}$ and $F_{\rm out}$  are the reflected and outward-radiated fluxes, and $\alpha$ is the inclination angle of the cloud (see Fig.~\ref{fig:scheme}). 
More precisely, the fluxes are the $EF_E$ quantities of the \cloudy\ output. 
Hereafter, we assume that all the radiation fields are given per logarithmic energy interval, unless otherwise stated. 
Attenuated source flux that also leaves the outer rim of the cloud is concentrated in a narrow solid angle that practically never contains the direction of interest and hence was ignored.

Then the clouds were treated as a continuous emitting medium with emission coefficient equal to the spatial density of the clouds $n_{\rm c} = 3f/4\pi R_c^3$ (where $f$ is the volume filling factor, i. e. fraction of the volume filled by the clouds) multiplied by the output of an individual cloud $dL_1/d\Omega$:
\begin{equation}\label{E:jgen}
\displaystyle j \!\!=\!\! n_{\rm c} \frac{dL_1}{d\Omega} \!\!=\!\!
\dfrac{3f}{4R_c} \!\! \left( \! \frac{1+\cos
  \alpha}{2} F_{\rm in}(R)\! + \! \frac{1-\cos \alpha}{2} F_{\rm out}(R)\! \right)\!.
\end{equation}
As the jets of blazars are directed close to the line of sight, we can restrict our analysis to the gamma-ray sources  and observers located along the axis of the BLR (see Fig.~\ref{fig:scheme}). 
The intensity observed from the viewpoint on the axis at the height of $H$ in the direction set by the polar angle cosine $\mu=\cos\theta$ equals
\begin{equation}\label{E:intty}
I(\mu, H) = \int_0^{+\infty} j(\mu, H, l) dl,
\end{equation}
where $l$ is the distance along the integration line. 
Both angle $\alpha$ and distance $R$ change along the line of integration and may be found using cosine theorems as $\displaystyle R=\sqrt{H^2+l^2+2Hl\cos\theta}$ and $\cos \alpha = (l^2+R^2-H^2)/2Rl$. 

\begin{figure}
 \centering
\includegraphics[width=\columnwidth]{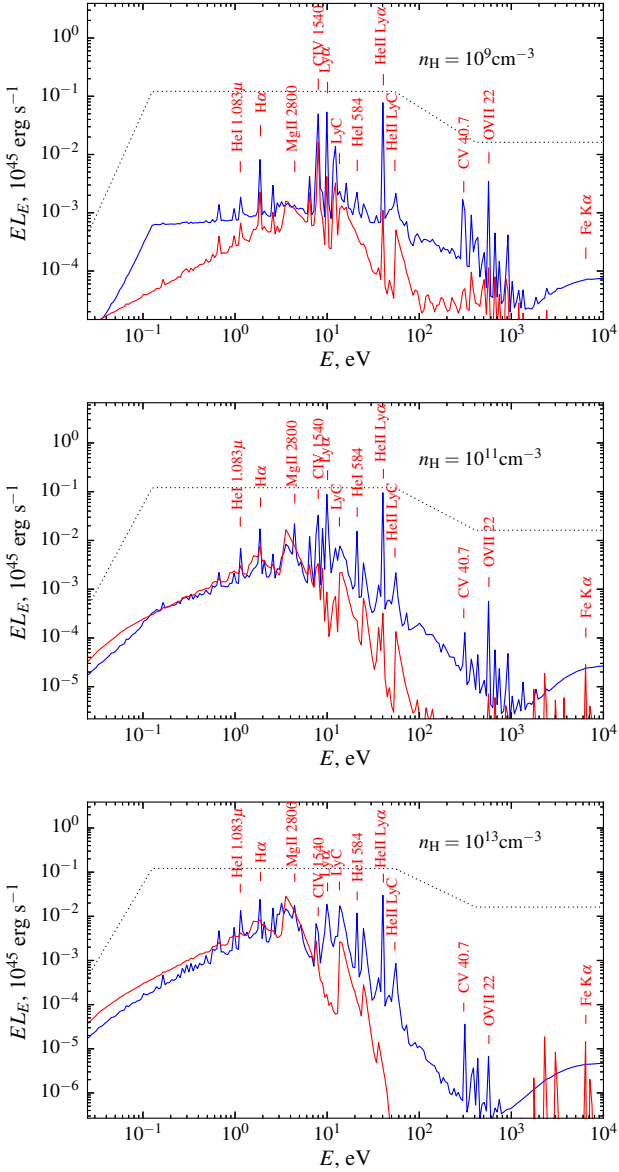}
\caption{ 
Integral spectra of the model BLR for the source luminosity 
$L=10^{45}\ergl$, $n_{\rm H} = 10^9$,
$10^{11}$,  and $ 10^{13}\cmc$ (from top to bottom),
$N_{\rm H} = 10^{23}\cmsq$, $R_{\min}=10^{-3}$ and $R_{\max} =1\pc$. Red and
blue lines show the contributions of the outward and inward-emitted parts of
the spectra. { Dotted black lines show the incident continuum. }
}
\label{fig:ltot}
\end{figure}

\begin{figure}
 \centering
\includegraphics[width=\columnwidth]{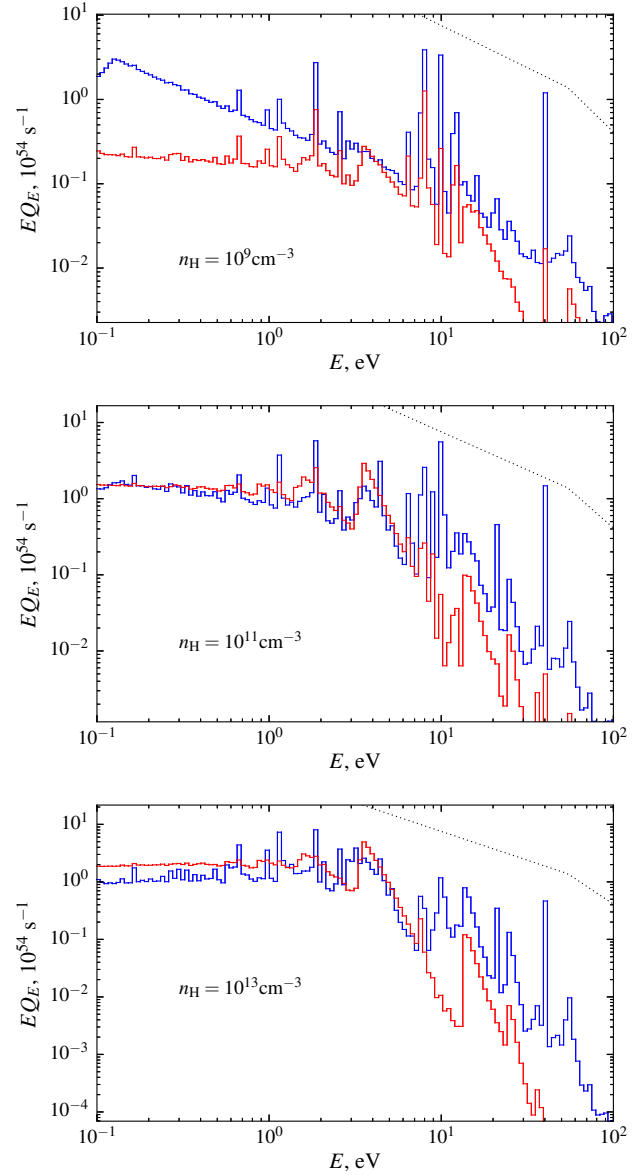}
\caption{ 
The same as Fig.~\ref{fig:ltot}, but photon production rates are shown instead of
luminosities. The energy resolution was intentionally degraded to about 5 per cent. 
}
\label{fig:qtot}
\end{figure}

We assumed that the clouds are distributed sparsely enough to neglect any shading so that the continuum shape is identical for all the BLR clouds. 
Probability for one cloud to appear between another and the source may be expressed in terms of the following ``overlap'' optical depth:
\begin{equation}
\tau_{\rm over} = n_{\rm c} \pi R_c^2 R_{\max} = \dfrac{3}{4} \dfrac{f R_{\max} }{R_c}.
\end{equation}
Our assumptions require a small value of $\tau_{\rm over} \ll 1 $. 
On the other hand, the total luminosity of BLRs is usually rather high (broad emission lines contain several per cent of the bolometric luminosity, significant part of
the UV continuum may be produced by emission reprocessed in BLR), hence the BLR should intercept significant part of the luminosity of the central source. 
For certainty, we hereafter assume a fixed total covering factor of the clouds $C = \tau_{\rm over} \cos \theta_1 = 0.1$. 
The filling factor depends on the geometry, $R_{\rm c}$ and  $C$: 
\begin{equation}\label{E:fgen}
f=\dfrac{4 R_{\rm c} C}{3R_{\max} \cos\theta_1} =\dfrac{4}{3} \dfrac{R_{\rm
    c}}{R_{\max} } \tau_{\rm over}.
\end{equation}
For $N_{\rm H}=10^{23}\cmsq$, $n_{\rm H}=10^{11}\cmc$, the size of a cloud $R_{\rm c} \simeq 1.6\times 10^{-7}\pc \simeq 7 R_\odot$ and the filling factor $f\sim 10^{-7}$ for $R_{\max} =0.1\pc$. 
The total number of clouds is $N_{\rm tot} \simeq \dfrac{4\pi}{3} R_{\max} ^3 n_{\rm c} \simeq \tau_{\rm over}^3 f^{-2}$. 
Varying $n_{\rm H}$ by a  factor of 100 from the fiducial value of $10^{11}\cmc$, the cloud radius would vary by the same amount and the filling factor then changes  also by about four orders of magnitude $f\sim 10^{-9}\div 10^{-5}$. 
As the likely overlap optical depth is $\tau_{\rm over} \sim 0.1\div 1$ (depending on BLR geometry), the number of clouds may vary between $10^7$ and $10^{18}$. 

\begin{figure*}
 \centering
\includegraphics[width=0.8\textwidth]{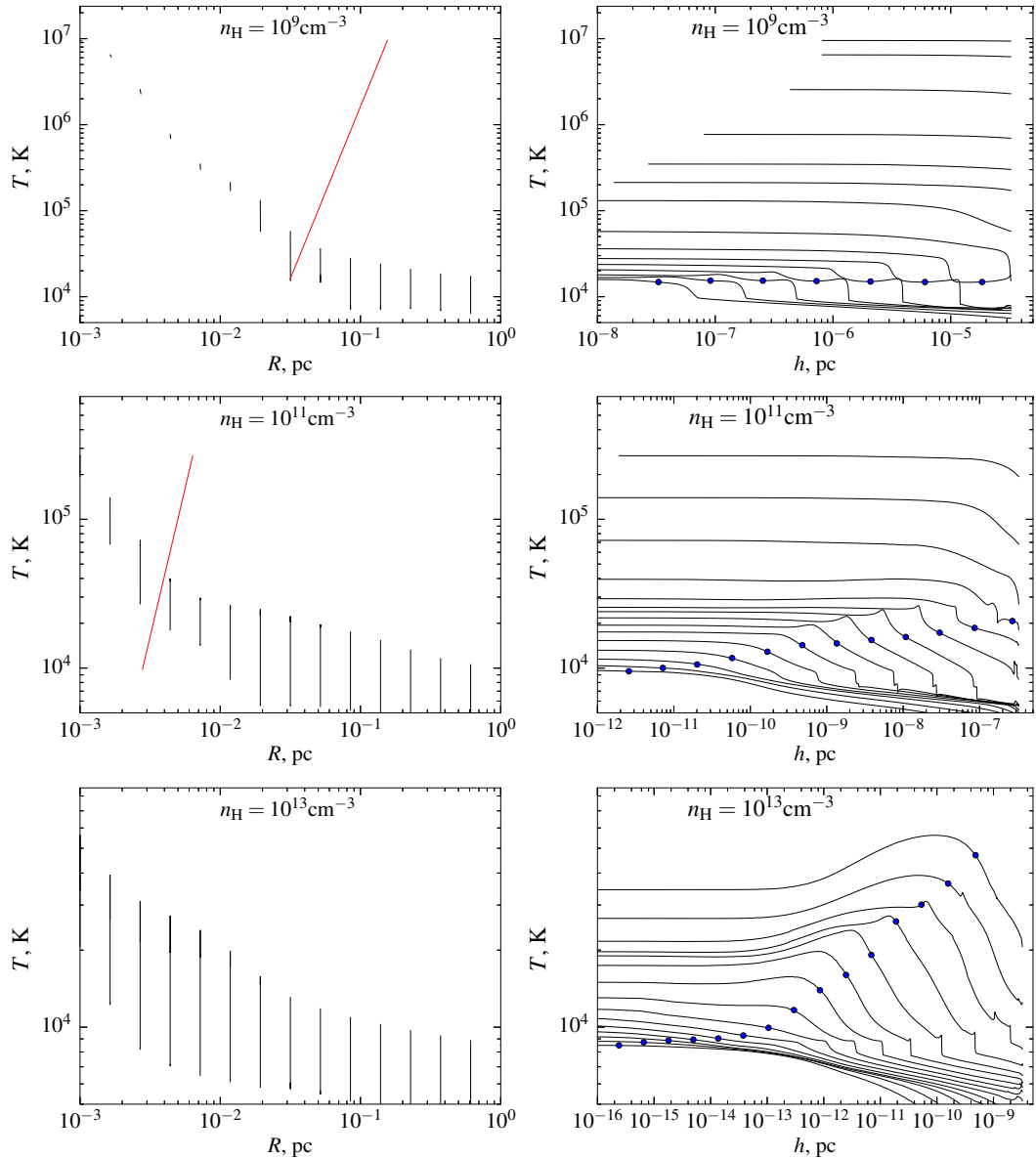}
\caption{ 
Temperature ranges and profiles for model clouds situated at different distances from the central ionizing source. 
In the left panels, temperature is shown as a function of distance to the source. 
The inclined red lines mark the region of complete ionization given by expression~(\ref{E:rion}). 
In the right panels, temperature dependence upon the thickness of the slab is shown. 
Ionization front positions as given by expression~(\ref{E:ithick}) are shown with small circles. 
}
\label{fig:tempss}
\end{figure*}

\subsection{Appearance of the BLR}\label{sec:blr}

Integrating emission coefficient allows to recover the integral spectrum and other properties of the BLR. 
These quantities may be compared with the observational data, both as a consistency check and as an indirect way to recover the morphology of the region. 
Fig.~\ref{fig:ltot} shows sample integrated spectra with the inward and outward contributions shown separately. 
The same is shown in Fig.~\ref{fig:qtot} in terms of the photon production rates. 
The spectra have different contributions, including free-free emission (smooth flat $L_E \simeq const$ continual emission below $\sim 1\eV$), recombination continua of hydrogen, \ion{He}{I} and \ion{He}{II} and strong emission lines of hydrogen, helium and heavy elements. 
In the extreme UV range above 10 eV, the strongest are the lines of helium-like carbon and oxygen. 
For outward emission from dense gas, the emission lines disappear, but recombination edges are always present in emission. 
Reflected spectra are profoundly different from the spectra produced by the outer sides of the emitting clouds, the difference increases with increasing $N_{\rm H}$. 
One evident difference is the reflected X-ray emission from the inner side that is at energies higher than several keV well approximated by the source spectrum
multiplied by the optical depth to Thomson scattering and the total covering factor $C\tau_{\rm sc} = C \sigma_{\rm T} N_{\rm H} \sim 0.0067C_{-1} N_{\rm H,23}$. 
At lower densities, reflection by free electrons is also visible at low energies. 
For large densities, the reflected component is underestimated by \cloudy\ that does not include the contribution of bound electrons to Thomson scattering (see Appendix~\ref{sec:app:xstar}). 

Depending on the ionizing flux, the model slab may be either totally ionized and thus matter-bound or may contain an ionization front and thus become ionization-bound. 
This is clearly illustrated by the temperature slices of different model slabs (see Fig.~\ref{fig:tempss}). 
The thickness $h$ of the ionized layer is determined by the balance of ionization and recombination processes as
\begin{equation}
q_{\rm i} = \alpha_{\rm i} n_e n_{\rm i} h,
\end{equation}
where $\alpha_{\rm i}$ is the recombination coefficient, weakly-dependent on temperature $T$, and $q_{\rm i}$ is the ionizing photon flux. 
For pure hydrogen plasma ionized by the EUV radiation, it is safe to assume $n_e = n_{\rm i} = n_{\rm H}$ and $\alpha_{\rm i} \simeq 2.6\times 10^{-13}\sqrt{T_4}\,{\rm cm}^{3} {\rm s}^{-1}$\citep{osterbrock}.
Thus, 
\begin{equation}\label{E:ithick}
\begin{array}{l}
h \simeq \dfrac{Q_{\rm ion}}{4\pi R^2 \alpha_{\rm i} n_{\rm H}^2} \simeq 10^{-11} \dfrac{Q_{\rm ion,55}T_4^{1/2}}{R_{\rm pc}^2 n_{\rm H,11}^2}\pc \\
\qquad{} \simeq 2.6\times 10^{-12}
\dfrac{L_{45}T_4^{1/2}}{R_{\rm pc}^2 n_{\rm H,11}^2}\pc,
\end{array}
\end{equation}
Inside a certain critical radius $R_{\rm ion}$, the thickness of the ionized
region is larger than the size of the cloud, $h>R_{\rm c}$,
\begin{equation}\label{E:rion}
\begin{array}{l}
R_{\rm ion} = \sqrt{\dfrac{Q_{\rm ion}} {4\pi \alpha N_{\rm H} n_{\rm H} } }
\simeq
5.6\times 10^{-3}\dfrac{Q_{\rm ion,55}^{1/2} T_4^{1/4}} {N_{\rm H,23}^{1/2} n_{\rm H,11}^{1/2}}\pc\\
\qquad{} \simeq 2.9\times 10^{-3}
\dfrac{L_{45}^{1/2} T_4^{1/4}} {N_{\rm H,23}^{1/2} n_{\rm H,11}^{1/2} } \pc.
\end{array}
\end{equation}
The temperature structures of model clouds are shown in Fig.~\ref{fig:tempss}. 
The size $R_{\rm ion}$ of the completely ionized region is proportional to the conventional BLR size that also scales with luminosity approximately as $R_{\rm BLR} \propto \sqrt{L}$. 
In particular, \citet{kaspi07} give an observational relation that may be written as 
\begin{equation}\label{E:Kaspi}
R_{\rm BLR} \simeq 9\times 10^{-3}L_{45}^{0.55\pm 0.05}\pc
\end{equation} 
that is very close to the size of the full-ionization region for $n_{\rm H} = 10^{10}\cmc$ and $N_{\rm H} = 10^{23}\cmsq$. 
To convert the normalization used by \citet{kaspi07} in equations (2) and (3) of their work, we assumed $\lambda L_\lambda \simeq 0.2 L_{\rm tot}$ between 0.12\eV\ and 4\Ry\ following from the spectral shape we use. 
This similarity suggests that the observational estimates of $N_{\rm H}$ may indeed reflect the column density of {\it ionized hydrogen} and thus be only lower estimates.

\begin{figure}
 \centering
\includegraphics[width=0.9\columnwidth]{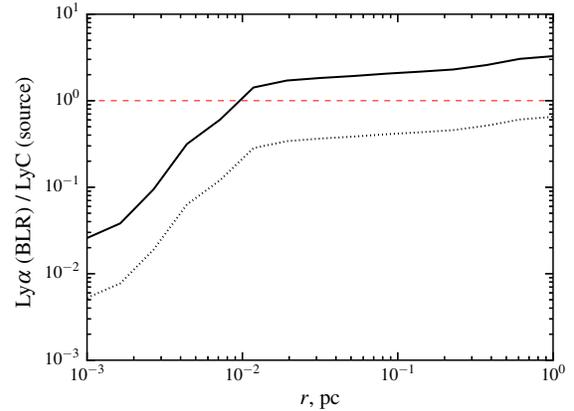}
\caption{ 
The fraction of ionizing quanta (solid) and the fraction of incident energy flux (dotted) converted to  Ly$\alpha$ as functions of distance from the photon source. 
Model parameters  $L=10^{45}\ergl$, $N_{\rm H} = 10^{23}\cmsq$, $n_{\rm H} = 10^{11}\cmc$. 
Red dashed horizontal line is unity. 
Note that a single ionizing photon can produce several Ly$\alpha$ quanta. 
}
\label{fig:lyrat}
\end{figure}

Reflected spectrum is bright in fluorescence lines including Ly$\alpha$, \ion{He}{I}$\lambda$584\AAA\ and usual BLR lines like \ion{C}{IV}$\lambda$1549 and \ion{C}{III]}$\lambda$1909. 
In the harder, unobservable EUV range, the most prominent emissions are \ion{He}{II} Ly$\alpha$ line at 228\AAA, Ly$\alpha$ lines of hydrogen-like CNO elements near 20\div30\AAA\ and the helium-like \ion{C}{V}$\lambda$40.7\AAA\ line, the analogue of \ion{He}{I}$\lambda$584\AAA. 
Luminosity in Ly$\alpha$ is comparable to the incident luminosity above the Lyman edge and exceeds considerably the emitted luminosity in Lyman continuum. 
Significant part of the ionizing spectrum comes well above the Lyman edge, where the ionization cross section is much smaller than that near the edge. 
At the same time, recombination to the ground level produces much softer quanta that become efficiently trapped inside the cloud, destroyed, and converted to Lyman emission lines and higher-order recombination continua. 
This explains the enhancement of the Ly$\alpha$ emission in comparison to LyC (see Appendix~\ref{sec:app:greenhouse} for more details). 
Ly$\alpha$ line works as one of the main coolants, especially for lower ionization parameters. 
The fractions of hydrogen-ionizing quanta and the hydrogen-ionizing luminosity converted to Ly$\alpha$ emission are shown in Fig.~\ref{fig:lyrat} as functions of distance for our model with $N_{\rm H} = 10^{23}\cmsq$ and $n_{\rm H} = 10^{11}\cmc$. 

In total, from 25 per cent (for $N_{\rm H} = 10^{23}\cmsq$) to about a half of the incident radiation is reflected by a BLR cloud. 
At the same time, the amount of radiation emitted outwards is generally less than 10 per cent, with larger values reached for higher densities. 
In Table~\ref{tab:albedo} we show the integral reflection ``albedo'' as a function of the column density and volume hydrogen density. 
Reflected Ly$\alpha$ emission in the units of intercepted broadband flux is also given. 
In the outward diffuse emission, Ly$\alpha$ emission is weaker and becomes an absorption for higher densities and column depths. 
 
 \begin{table}\centering
\caption{Reflected and outward-emitted flux fractions for different BLR models. 
  Broad-band emission is integrated between 0.1 and 30~eV. 
  The fraction of the incident broadband flux emitted in Ly$\alpha$ inwards  is also given. 
  For all the models, the range of radii was $10^{-3}$ to $0.1$\pc.}
\label{tab:albedo}
\begin{tabular}{lcccl}
\hline
                     &                     & \multicolumn{2}{c}{broadband}\\
$N_{\rm H}$ ($10^{23}$\cmsq) & $n_{\rm H}$ ($10^{11}$\cmc) &  in & out & Ly$\alpha$\\
\hline
0.1 & 1 & 0.22 & 0.08 & 0.037 \\
1 & 0.01 & 0.16 & 0.04 & 0.07 \\
1 & 0.1 & 0.25 & 0.08 & 0.11 \\ 
1 & 1 & 0.28 & 0.12 & 0.11 \\
1 & 100 & 0.24 & 0.18 & 0.026 \\ 
10 & 0.1 & 0.39 & 0.10 & 0.10 \\ 
10 & 1 & 0.35 & 0.15 & 0.12 \\
10 & 10 & 0.31 & 0.18 & 0.075 \\
100 & 0.1 & 0.47 & 0.07 & 0.14 \\
100 & 1 & 0.36 & 0.13 & 0.054 \\
1000 & 0.1 & 0.48 & 0.07 & 0.14 \\
\hline
\end{tabular}
\end{table}

Ly$\alpha$ is surrounded by other strong emission features, including other Lyman lines and Lyman recombination continuum, \ion{C}{IV}$\lambda$1549 and other lines, together providing a strong contribution to the overall number at soft photons roughly between 5 and 20\eV\ (\ion{He}{I}$\lambda$584 may be considered the hardest line of the blend). 
The second very important emission, \ion{He}{II}\,Ly$\alpha$, is relatively isolated and thus is capable, as we will see in section~\ref{sec:tau}, of creating a sharp break in the gamma-ray absorption spectra.

\begin{table*}\centering
\caption{Line fluxes in Ly$\alpha$ units for different models integrated between $10^{-3}$ and $0.1$\pc. 
  Observational values were taken from the compilation
  of~\citet{baldwin95}. Estimated Ly$\alpha$ flux does not include the strong
wings beyond $\pm 0.5\eV$ from the core of the line. }
\label{tab:rlines}
\begin{tabular}{lc cccccc}
\hline
\multicolumn{2}{c}{Model} & \multicolumn{6}{c}{Lines}\\
$N_{\rm H}$, $10^{23}$\cmsq & $n_{\rm H}$, $10^{11}$\cmc & 
\ion{He}{II}\,Ly$\alpha$ & \ion{He}{I}$\lambda$584 &
\ion{C}{IV}$\lambda$1549 &
\ion{He}{II}$\lambda$1640+\ion{O}{III}$\lambda$1666 & \ion{C}{III]}$\lambda$1909 & \ion{Mg}{II}$\lambda$2800\\
\hline
 0.1 & 1 &  0.5 & 0.2 &
  0.25  & 0.17 & 0.5 & 0.6 \\ 
 1  &  0.01 & 0.8 & 0.02 &
0.5 & 0.09 & 0.03 & 0.0\\ 
 1  &  0.1 & 0.7 & 0.05 &
0.6 & 0.1 & 0.06 & 0.0\\ 
 1  &  1 & 0.6 & 0.12 &
0.2 & 0.09 & 0.11 & 0.11\\ 
 1  &  100 & 1.5 & 0.75 &
0.19 & 0.3 & 0.0 & 0.4\\ 
 10  &  0.1 & 0.6 & 0.05 &
0.6 & 0.12 & 0.07 & 0.0\\ 
 10  &  1 & 0.6 & 0.15 &
0.4 & 0.10 & 0.03 & 0.02\\ 
 10  &  10 & 0.9 & 0.4 &
0.2 & 0.12 & 0.015 & 0.09\\ 
 100  &  0.1 & 0.6 & 0.05 &
0.6 & 0.14 & 0.08 & 0.01\\ 
 100  &  1  & 0.6 & 0.15 &
 0.5 & 0.10 & 0.04 & 0.02 \\ 
1000 & 0.1 & 0.6 & 0.05 &
0.6 &  0.15 & 0.08 & 0.01 \\ 
\hline
\multicolumn{2}{c}{observed}
& -- & -- &  0.4--0.6 & 0.02--0.2 & 0.15--0.3 & 0.15--0.3 \\
\hline
\end{tabular}
\end{table*}

\begin{figure}
 \centering
\includegraphics[width=0.9\columnwidth]{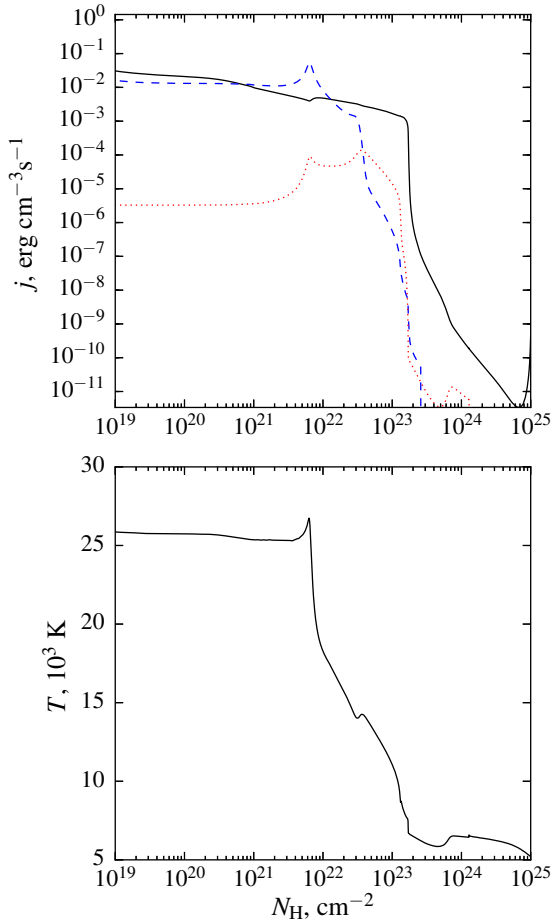}
\caption{ 
Top panel: Line emissivities as a function of the cumulative hydrogen column density in a cloud at distance $r=0.01\pc$. 
The solid black line corresponds to Ly$\alpha$, the dashed blue line is for \ion{C}{IV}$\lambda$1549, and the dotted red is for 
\ion{C}{III]}$\lambda$1909. 
Bottom panel:  The local temperature as a  function of the cumulative $N_{\rm H}$. 
Hydrogen ionization fraction is close to unity for $N_{\rm H} \lesssim 10^{24}\cmsq$. 
Here we took density of $n_{\rm H} =10^{10}\cmc$.  
}
\label{fig:linetrace}
\end{figure}

So far, the model we use has a large number of free parameters. 
Most of them may be constrained by the observational data. 
For instance, the radial sizes of BLRs are constrained observationally through reverberation studies. 
In particular, equation~(\ref{E:Kaspi}) allows to estimate the mean effective
radius of a BLR as $R\sim 0.01 \sqrt{L_{45}} \pc$. 
This estimate, however, refers to the variable component of the broad line emission. 
Variable emission of the more distant parts of the BLR is smeared by light propagation effects.
In different emission lines, the effective BLR sizes differ by about an order
of magnitude \citep{PW99}, hence, a large span of radii is required to
reproduce the radiation field of a BLR. 
Hereafter, we use the range $r=10^{-3}\div 0.1\pc$. 

At lower radii, the clouds are completely ionized, emit very little in
emission lines and contribute mainly through electron scattering. 
At larger radii, there is increasing contribution from intermediate ionization
species like \ion{C}{IV} and \ion{He}{II}. Lyman series lines dominate at the
largest distances we consider.
In Table~\ref{tab:rlines}, we compare the estimated line flux ratios to the
fluxes from the compilation of \citet{baldwin95}. 
All the fluxes are expressed in the units of Ly$\alpha$ flux.  

Independently of the range of radii used, it is difficult to match all the line ratios. 
In particular, lower-ionization lines like \ion{Mg}{II}$\lambda$2800 require systematically lower ionization parameters (and thus larger radii or larger volume and column densities) than higher-ionization lines like \ion{C}{IV}$\lambda$1549 and \ion{He}{II}$\lambda$1640. 
As it was mentioned in  numerous works \citep{KG00,negrete14,ruff12,moloney}, any reasonable fit aiming to reproduce the spectrum of a BLR in details should be multi-component. 
It is also instructive in more comprehensive models to consider more sophisticated radial distributions of cloud density. 

Another poorly-constrained quantity is the thickness of the cloud. 
Larger column densities $N_{\rm H} \gtrsim 10^{25}\cmsq$ seem to provide better results. 
Thin clouds are unable to reproduce some of the lines like \ion{C}{III]}$\lambda$1909 that form more efficiently in hot ionized gas but require attenuated EUV continuum.
In Fig.~\ref{fig:linetrace}, we show the emissivities in different emission lines as functions of depth. 
Evidently, to make a strong \ion{C}{III]} emission line, the cloud should be either relatively thick or shielded by other clouds. 
Spikes and breaks at $N_{\rm H}\sim 10^{23}$, $3\times 10^{23}$ and $8\times 10^{23}\cmsq$ are connected to changes in dominating ionization states of carbon. 
Some carbon lines have important contribution to cooling function that makes the local temperature sensitive to the distribution of carbon ionization states. 
The Ly$\alpha$ upturn at the large depth appears as a result of formulation of the escape probability  that is included in definition of emissivity used by \cloudy. 
As Ly$\alpha$ is an optically thick line, escape probability factor creates a considerable depression in the emergent emissivity. 

For high ionization parameters, $U\sim 1$, the cloud should be thicker than $10^{23}\cmsq$ to emit significantly in \ion{C}{III]}$\lambda$1909. 
The effects of higher column density are, however, indistinguishable from cloud overlap that we do not include in our model directly: $N_{\rm H, eff} \simeq  \tau_{\rm over } N_{\rm H}$.
There is evidence \citep{GaskellBC} coming from hydrogen line ratios that the column densities of individual clouds are in fact much smaller than the canonical $10^{23}\cmsq$, but the number of clouds is larger and effects of self-shading are important. 

\section{$\gamma$-ray propagation through the BLR radiation field}\label{sec:tau}

Jets in blazars shine at very small angles with the line of sight, hence we can estimate the optical depths for a gamma-ray photon moving along the symmetry axis away from the centre. 
In each point situated at a distance $H$ along the jet, we calculate the radiation field created by the BLR. 
The intensity for given $H$ and parameters of the model, coming from any given direction, was calculated according to the formulae of Section~\ref{sec:model:setup}. 
Knowing intensity distribution in a given point, it is possible to calculate the local opacity to pair production. 

Following \citet{QED}, \citet{nikishov} and \citet{gould}, we express the photon-photon absorption cross-section as
\begin{eqnarray}
\sigma_{\gamma\gamma} (s) & = & \displaystyle \frac{3}{8} \frac{\sigma_{\rm T}}{s^3}
\left[\left(s^2+s-\frac{1}{2}\right) \ln \left( \frac{\sqrt{s}+\sqrt{s-1}}{\sqrt{s}-\sqrt{s-1}}\right) \right. \nonumber\\
& - & \left.  \sqrt{s(s-1)} (s+1)\right], 
\end{eqnarray}
where  
\begin{equation}
s = \dfrac{E E^\prime (1-\cos \theta_{\gamma\gamma})}{2 (m_e c^2)^2}
\end{equation}
is the measure of centre-of-mass square energy (in units of the electron rest energy $m_ec^2$) of the pair created by the absorption process, $E$ and $E^\prime$ are the energies of the two photons (below, $E^\prime$ will be used for the energy of the soft photon), and $\theta_{\gamma\gamma}$ is the angle between their propagation  directions (for gamma-ray photons moving upward along the axis, $\cos \theta_{\gamma\gamma} = -\mu = -\cos \theta$). 
Cross-section is zero for $s\leq 1$ where conservation laws do not allow the two given photons to create an electron-positron pair. 
Hence any given spectral detail located at some $E^\prime$ in the soft spectrum is able to interact with gamma-ray photons with energy $E$ above the threshold energy 
\begin{equation}
E_{\rm thr} = \dfrac{(m_ec^2)^2}{E^\prime} \simeq 26 \dfrac{10\eV}{E^\prime} \GeV,
\end{equation}
which is the minimal energy for absorption in a head-on collision. 
The maximal cross-section of about $0.25\sigma_{\rm T}$ is reached at the energy about two times larger.  
It will be further shifted towards higher energies if all the soft photons arrive at large impact angles, $\theta \gtrsim  \pi/2$, as it is expected if the BLR is flattened or if the gamma-ray emission is generated at distances larger than the size of the BLR. 
The observed break in the gamma-ray spectrum will shift to the effective threshold energy of
\begin{equation}\label{E:eest}
E_{\rm eff} \sim \dfrac{2}{1+ \mu_{\max} } E_{\rm thr}, 
\end{equation}
where $\mu_{\max}= -\cos \theta_{\gamma\gamma, \min}$ corresponds to the minimum interaction angle between the photons. 
When the distance from the BLR greatly exceeds its size, $H\gg R_{\max} $, $1+ \mu_{\max}$ approaches zero,  $E_{\rm eff} \sim 4 \left({H}/{R_{\max} }\right)^2 E_{\rm thr}$, and all the opacity breaks become increasingly smeared and blueshifted. 
As we will see below, the optical depth also rapidly goes down in this case.
Below in Section~\ref{sec:res}, this is confirmed by numerical calculations. 

The photon-photon absorption coefficient for gamma-ray quanta is
\begin{equation}
\alpha_{\gamma\gamma}(E)\! \! = \! \! \dfrac{2\pi}{c}\! \! \int_{-1}^{1} \! \! \! \! d\mu \int_{0}^{+\infty} \! \! \! \!\! \!
(1+\mu)\sigma_{\gamma\gamma}(s) N_{E^\prime}(\mu)dE^\prime ,
\end{equation}
where $N_E$ is the differential photon number flux and the factor $2\pi$ accounts for azimuthal angle range. 
The photon number flux 
\begin{equation}
N_E = \dfrac{E I_E}{E^2},
\end{equation}
where $E I_E$ intensity is given by equations~(\ref{E:jgen}) and (\ref{E:intty}). 

Optical depth from the $\gamma$-ray source towards the observer is then expressed as the integral along the line of sight
\begin{eqnarray}\label{E:tau}
\tau_{\gamma\gamma}(E,H) & = & \! \! \! \! \int_H^{+\infty} \alpha_{\gamma\gamma} dh =
 \pi \frac{E}{c (m_e  c^2)^2} \int_{H}^{+\infty} dh  \nonumber \\
& \times&\! \! \! \!   \int_{-1}^{1} (1+\mu)^2 d\mu \int_1^{+\infty}
\frac{\sigma_{\gamma\gamma}(s)}{s^2} E^\prime I_{E^\prime }ds, 
\end{eqnarray}
where energy $E^\prime$ is expressed through the integration variable $s$ as:
\begin{equation}
E^\prime = \dfrac{2 (m_e c^2)^2}{(1+\mu) E} s.
\end{equation}

\subsection{Size-luminosity degeneracy}\label{sec:scale}

In our study, we used a fixed value of luminosity and a fixed shape of ionizing continuum. 
While the shape of the continuum is possibly important even in the far-infrared part (see \citealt{anisofer}), the luminosity is degenerate with the spatial scales. 
The local emission coefficient of BLR clouds depends upon the flux density rather than luminosity (see equation~(\ref{E:jgen})) but also scales inversely with the size of the BLR $j \propto L R^{-2} R_{\max} ^{-1}$. 
The scaling with $R_{\max} $ is a consequence of the fixed covering factor assumption we use in our study: in equation~(\ref{E:jgen}), the $f/R_{\rm c}$ multiplier is proportional to $\tau_{\rm over}/R_{\max} $ (see expression~(\ref{E:fgen})).
Expression~(\ref{E:tau}) contains two multipliers scaling with the distances, height $H$ and length $l$, hence the observed optical depth is proportional to the luminosity over the size $\tau_{\gamma\gamma} \propto L/R$ as long as $H\propto R$. In general, the optical depth is the same for identical $L/R$ and $H/R$. 
As we have seen above, different radius estimates for BLR scale with the square root of luminosity $R\propto \sqrt{L}$, that means $\tau \propto \sqrt{L}$. 
For instance, the maximal possible optical depth should scale as $\sqrt{L}$. 

\section{Results}\label{sec:res}

\subsection{Thin sphere case}\label{sec:res:sph}

In this case ($\theta_1 =0$, $R_{\max} \simeq R_{\min}$), one should expect a strong and rapid change in radiation fields near the surface of the sphere. 
\citet{tavecchio12} considered a similar problem and obtained an abrupt opacity drop near the surface of the sphere greatly exceeding the factor of several that may be attributed to geometrical reasons. 
It is unclear if the authors considered the BLR clouds emitting isotropically or not. In any case, there is an additional drop in opacity at the shell radius, especially at lower energies. 

\begin{figure}
 \centering
\includegraphics[width=0.8\columnwidth]{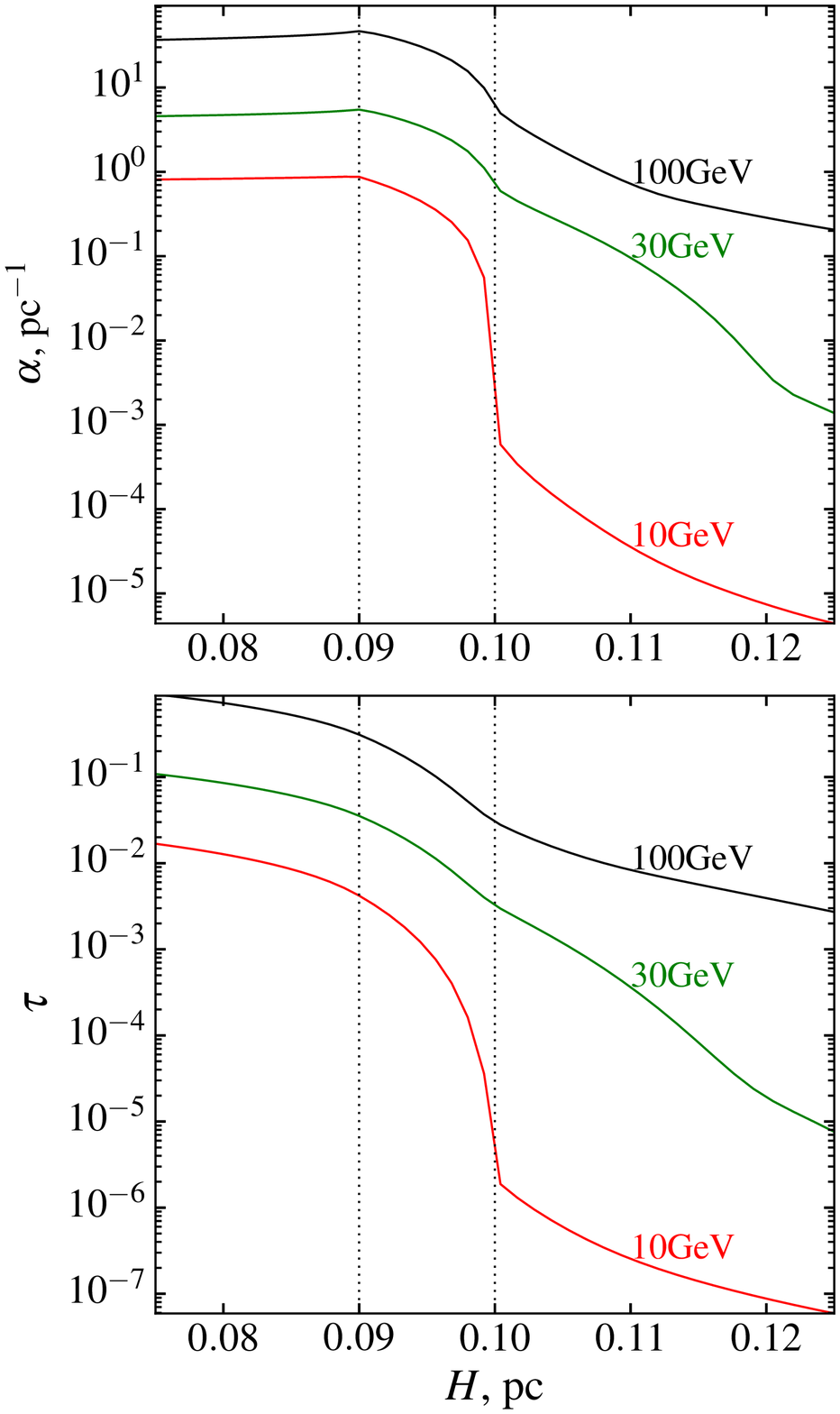}
\caption{Opacities $\alpha_{\gamma\gamma}$ and optical depths $\tau_{\gamma\gamma}$ as functions of distance from the central source at energies 10, 30 and 100~GeV (red, green and black lines, respectively) for the case of a thin spherical shell BLR with $0.09<R<0.1\pc$, $n_{\rm H} = 10^{11}\cmc$, $N_{\rm H} =
10^{23}\cmsq$. 
The vertical dotted lines mark the boundaries of the shell. 
}
\label{fig:sing2}
\end{figure}

\begin{figure}
 \centering
\includegraphics[width=0.6\columnwidth]{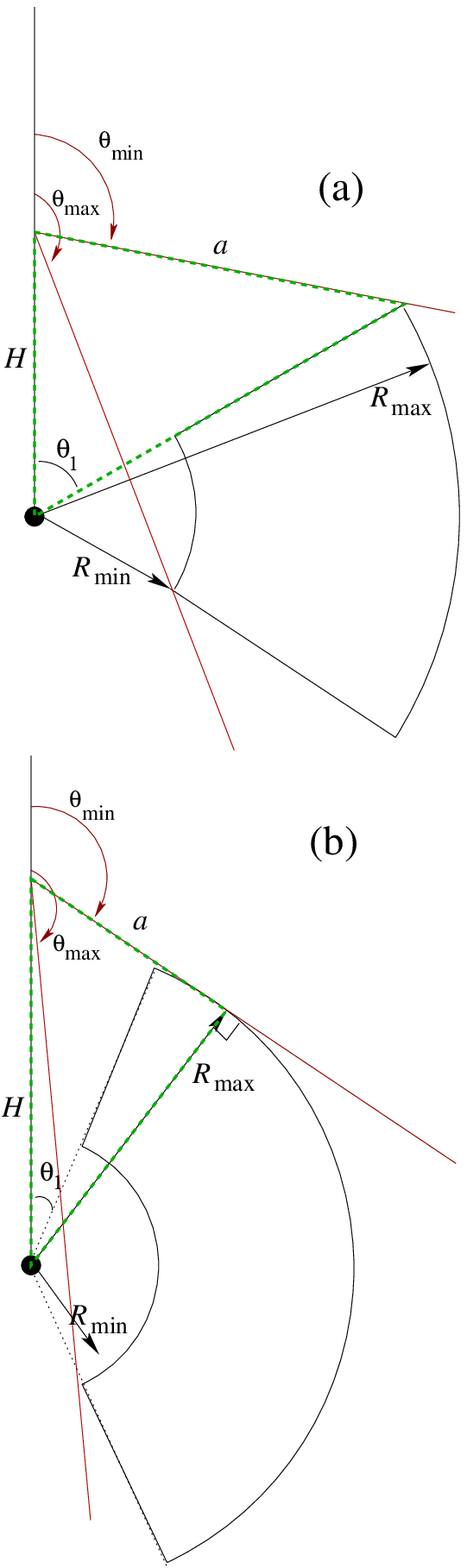}
\caption{Sketches illustrating the calculation of maximal possible impact angle cosine $\mu$: (a) for $H<R_{\max} /\cos\theta_1$ and (b) $H>R_{\max} /\cos\theta_1$. 
Red lines show the extreme lines of integration. 
Solid black lines show the boundary of the BLR and the axis of the jet, as in Fig.~\ref{fig:scheme}.
}
\label{fig:sub}
\end{figure}

We confirm this effect (see Fig.~\ref{fig:sing2}) and find it practically unimportant for larger energies $E\gtrsim 100\GeV$ but increasingly strong below $\sim 30\GeV$. 
Energy dependence suggests that the reason for the rapid decline of opacity is in coupling between the energy and incidence angle. 
At any given energy $E$, only photons with energies $E^\prime \gtrsim {(m_ec^2)^2}/{ (1+\mu)E}$ can contribute, and the rapid change of angular distribution of radiation near the surface of the sphere leads to a rapid drop in opacity, especially near the opacity threshold energy. 
In particular, at 30\GeV, Ly$\alpha$ creates considerably large opacity for $\mu>0$ (head-on collisions) but has zero contribution for $\mu<0$. 
Hence, the relatively large drop at 20$\GeV$ may be easily explained by Ly$\alpha$ emission plus the overall rapid decrease in photon number with energy at 10$\div$20\eV\ (see Fig.~\ref{fig:qtot}). 

\begin{figure}
 \centering
\includegraphics[width=0.9\columnwidth]{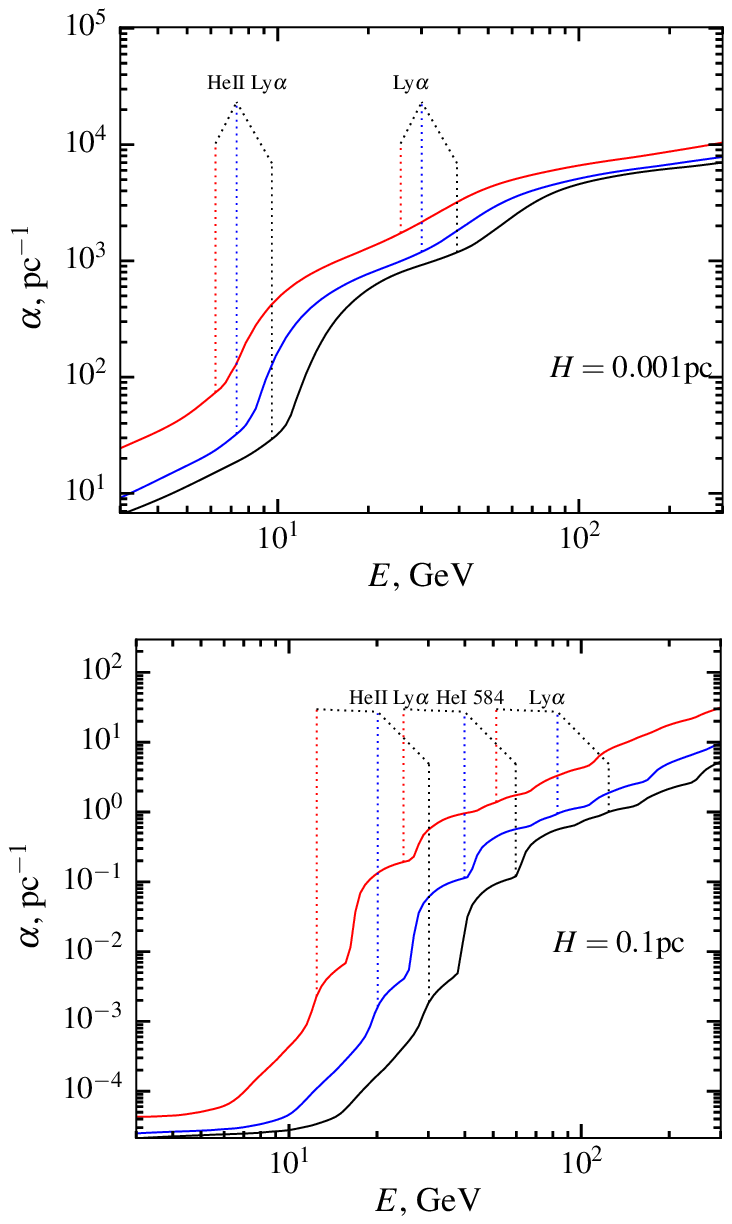}
\caption{ 
Opacity coefficients $\alpha_{\gamma\gamma}$ to pair production at the heights of $H=10^{-3}$ (upper panel) and $0.1$\pc\ (lower panel). 
Here we assume the outer radius of $R_{\max} =0.1$\pc, $n_{\rm H} = 10^{11}\cmc$, and $N_{\rm H}=10^{23}\cmsq$ and consider different geometries:
spherical (upper red  curve), intermediate ($\theta_1=45^\circ$, middle blue  curve) and disc-like ($\theta_1=72^\circ$, lower black  curve). 
Dotted vertical lines show the positions  of the spectral breaks corresponding to H, \ion{He}{I}$\lambda$584\AAA\ and \ion{He}{II} Ly$\alpha$ lines predicted by equations~(\ref{E:emum}), (\ref{E:mum1}), and (\ref{E:mum2}). 
}
\label{fig:tauar}
\end{figure}

\begin{figure*}
 \centering
\includegraphics[width=0.75\textwidth]{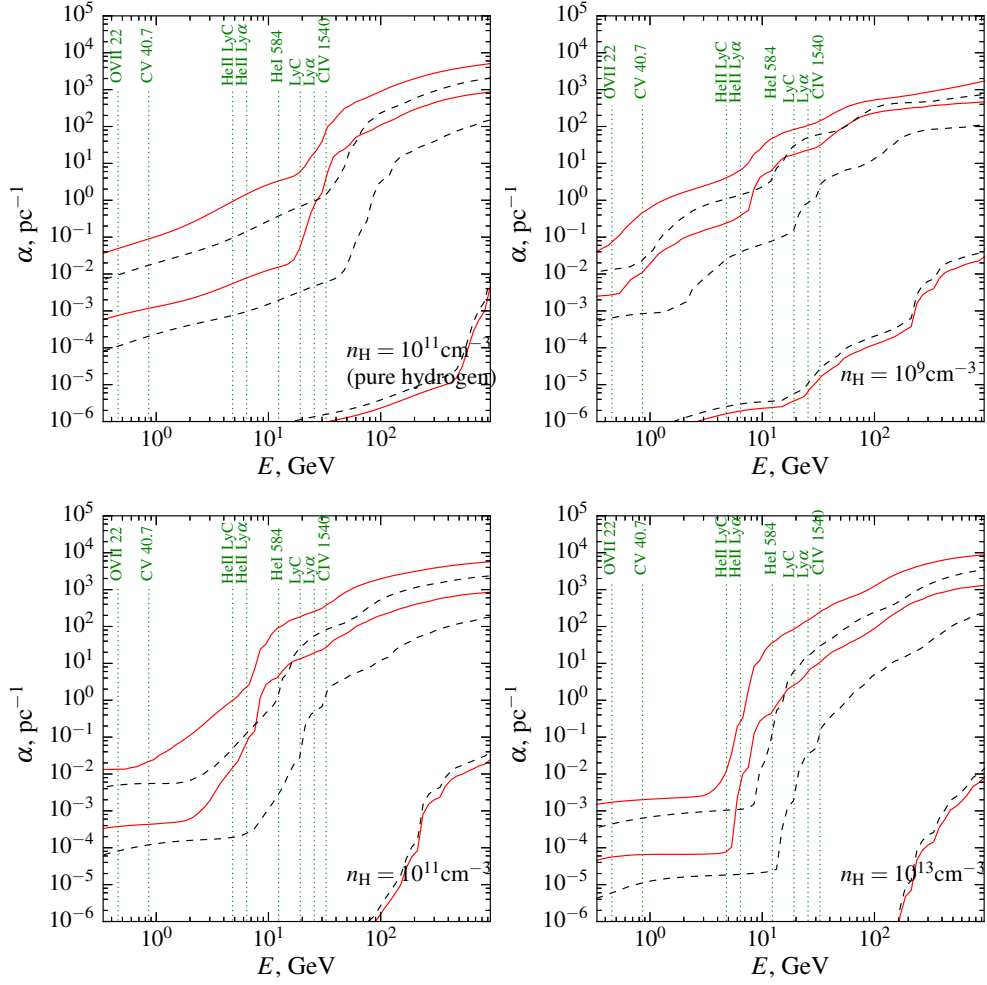}
\caption{Opacity coefficients $\alpha_{\gamma\gamma}$ for photon-photon pair production for three different distances from the central source ($0.01$, $0.05$ and $0.3$\pc; lines from top to bottom), four different BLR models with different volume densities, and two geometries: spherical (upper red lines) and disc-like with $\theta_1=72\degr$ (lower black dashed lines). 
Column density is set to $N_{\rm H} =10^{23}\cmsq$, $R_{\max} = 0.1\pc$.
}
\label{fig:smaps}
\end{figure*}

\begin{figure*}
 \centering
\includegraphics[width=0.75\textwidth]{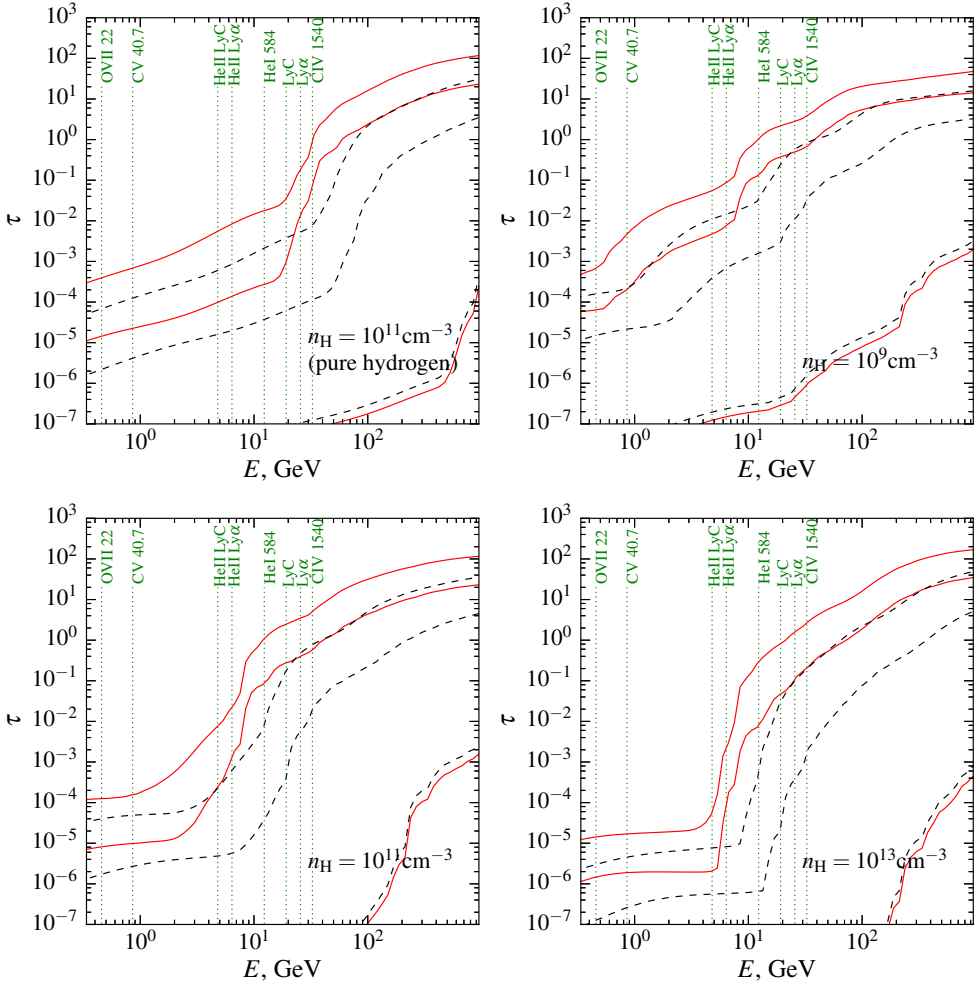}
\caption{ 
Optical depths for photon-photon pair production towards infinity for the models shown in Fig.~\ref{fig:smaps}. 
}
\label{fig:taus}
\end{figure*}

\subsection{The role of geometry}\label{sec:res:geom}

One of the main goals of the present work is to estimate the dependence of gamma-ray absorption upon the shape of the BLR. 
As it is evident from Section~\ref{sec:tau}, geometrical effects are tightly coupled to energy shifts. 
All the spectral details shift with the cosine of incidence angle as $E\propto \left(1+\mu\right)^{-1}$. 
The soft photons of energy $E^\prime$ closest to head-on collision create a break at
\begin{equation}\label{E:emum}
E_{\min} = \dfrac{2(m_ec^2)^2}{(1+\mu_{\max} )E^\prime},
\end{equation}
where $\mu_{\max} =\cos\theta_{\min}$ is the maximal possible $\cos
\theta$. 
Depending upon the ratio $H/R_{\max} $ and the opening angle $\theta_1$, $\mu_{\max} $ may correspond either to the visible rim of the upper conical surface of the BLR (see Fig.~\ref{fig:sub}a) or to the tangency point between the line of integration and the outer spherical surface of the BLR (Fig.~\ref{fig:sub}b). 
In the first case, $\mu_{\max} $ is most convenient to obtain using cosine theorems 
\begin{equation}\label{E:cos1}
a^2=H^2+R_{\max} ^2 - 2 H R_{\max} \cos \theta_1
\end{equation}
(where $a$ is the distance to the outer boundary edge) and 
\begin{equation}\label{E:cos2}
R_{\max} ^2=a^2+H^2 + 2 a H  \mu_{\max} , 
\end{equation}
which may be combined to obtain an expression for $\mu_{\max} $. 
For $H<R_{\max} /\cos \theta_1$, we get 
\begin{equation}\label{E:mum1}
\mu_{\rm max,1} =\displaystyle \frac{1}{a}\left( R_{\max} \cos \theta_1 -
  H\right),
\end{equation}
while for $H>R_{\max} /\cos \theta_1$ (see Fig.~\ref{fig:sub}b), the maximal possible $\mu$ corresponds to the line of integration being a tangent to the outer spherical surface limiting the BLR and
\begin{equation}\label{E:mum2}
\mu_{\rm max,2} =- \displaystyle \sqrt{1-\dfrac{R_{\max} ^2}{H^2}}.
\end{equation}
In Fig.~\ref{fig:tauar}, we show an example illustrating the shift of spectral details with change of geometry for a realistic model with bright Ly$\alpha$ and \ion{He}{II}\,Ly$\alpha$ emissions. 
Usually, there are visible spectral breaks at the minimal energies corresponding to these two emission features. 
At larger distances $H\sim 0.1\pc$, contribution of numerous other lines becomes important, and the features corresponding to the lines formed at $R\ll H$ become smeared.
Spectral breaks become sharper as the geometry becomes more and more oblate. 
This effect is the strongest for $H\ll R_{\max} $, when disc-like geometry means that the incidence angles concentrate at $\mu\simeq 0$. 
The strongest emission lines are emitted up to the maximal radius, hence the positions of the breaks are determined mainly by $\mu_{\max} $, but may be shifted to higher energies for the spectral details emitted mainly in the inner parts of the BLR. 
 
To illustrate the effects of absorption variations with geometry, we have calculated two sets of models: one for the spherically symmetric distribution of BLR clouds, and another for a strongly flattened distribution with $\theta_1 = \pi/2 - \pi/10 =72^\circ$. 
In Fig.~\ref{fig:smaps}, we show gamma-ray opacity coefficients in the GeV range for different distances along the jet. 
Three different distances, 0.01, 0.05 and 0.3\pc, are shown. 
Closer to the central source the radiation field does not change strongly, while at larger distances, the radiation field decays very rapidly. 
Positions of the spectral breaks scale approximately with $1/(1+\mu_{\max})$, hence the opacities are nearly identical functions of $E(1+\mu_{\max})$. 
This implies that small variations of the BLR geometry lead to huge
  changes in $\tau_{\gamma\gamma}$ if the opacity at a considered energy has a
  large derivative. 

Integrating the opacities to infinity allows us to calculate the total optical depths, shown in Fig.~\ref{fig:taus}. 
Two-three or more breaks are distinguishable in most of the spectra, the strongest of them caused by hydrogen and helium Ly$\alpha$+LyC emissions, \ion{He}{I}$\lambda$584 line, and Balmer continuum dominating the emission at about $3\div5\eV$. 
With increasing $\theta_1$, all the optical depths shift to higher energies in accordance with expressions (\ref{E:emum}), (\ref{E:mum1}) and (\ref{E:mum2}).

\subsection{Effects of the gas density }\label{sec:res:phys}
 
\begin{figure}
 \centering
\includegraphics[width=0.8\columnwidth]{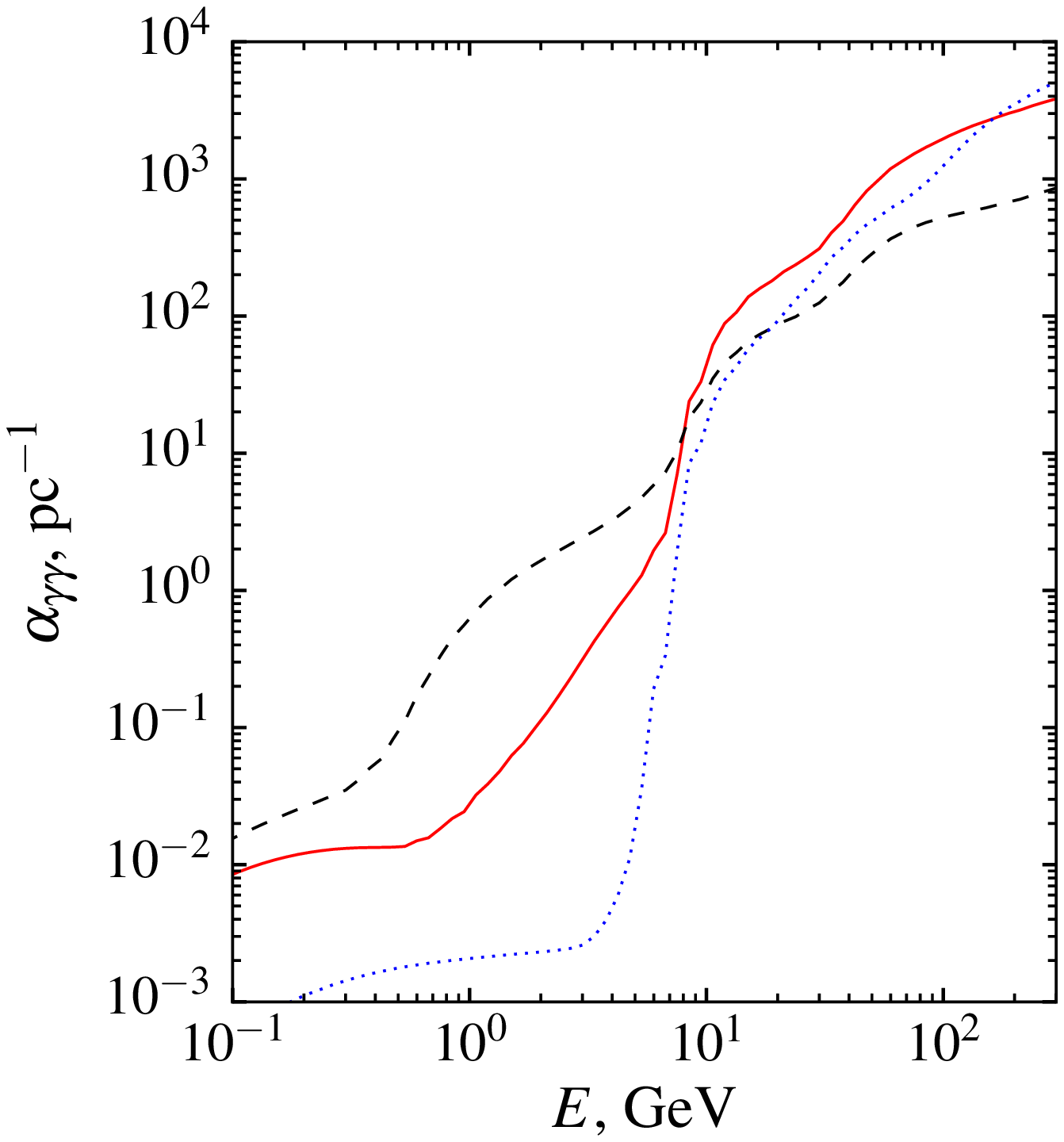}
\caption{Effects of hydrogen density upon the opacity for a fixed distance
  $H=0.01\pc$, $N_{\rm H}=10^{23}\cmsq$, geometry is spherical. 
  Black dashed, red solid and blue dotted curves correspond to $n_{\rm H}=10^9$, $10^{11}$ and  $10^{13}\cmc$, respectively. 
}
\label{fig:algoden}
\end{figure}

Volume gas density affects the spectrum of the BLR in several ways. 
First, the line-to-continuum contrast is stronger for a more rarefied gas (see Fig.~\ref{fig:ltot}). 
Next, the shape of the continuum changes: at the lowest densities and largest ionization parameters, it is dominated by reflected ionizing continuum, then the contribution of free-free and free-bound processes becomes more important, and then, the radiation gets thermalized. 
All this happens at different column densities for different densities and energies, but all the stages are recognizable in Fig.~\ref{fig:ltot}. 
The peaking continuum at large densities ($n_{\rm H}\gtrsim 10^{11}\cmc$) leads to a very steep slope in the region of $(5\div 10)/(1+\mu_{\max} )\GeV$ associated with the region above and around Ly$\alpha$. 
The impact of the soft X-ray lines (helium-like carbon and oxygen) becomes negligible at higher densities. 
For a fixed distance from the central source, the dependence upon hydrogen density in shown in Fig.~\ref{fig:algoden}.
 
\begin{figure}
 \centering
\includegraphics[width=0.8\columnwidth]{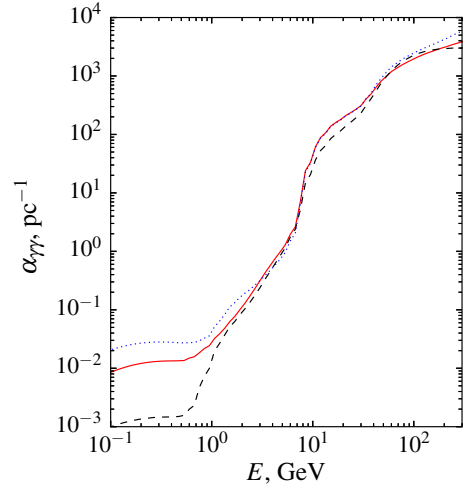}
\caption{Effects of hydrogen column density upon the opacity  coefficients. 
The density was set to $n_{\rm H}=10^{11}\cmc$, geometry is
  spherical, distance $H=0.01\pc$. 
  Black dashed, red solid and blue dotted curves correspond to $N_{\rm H}=10^{22}$, $10^{23}$ and $10^{25}\cmsq$, respectively.  
}
\label{fig:algocol}
\end{figure}

The impact of column density is much smaller. For $N_H\lesssim 10^{22}\cmsq$, the emissivity and, subsequently, the pair-production optical depth  scale approximately as $\propto \sqrt{N_{\rm H}}$. 
Above $10^{23}\cmsq$, spectral shape and intensity in the range $3\div 30\eV$ saturate, and the optical depths between 1 and 30\GeV\ appear constant to accuracy about 20 per
cent. 
However, the opacities at lower and higher energies proceed growing with column density (see Fig.~\ref{fig:algocol}).

\subsection{Anisotropy effect}

\begin{figure}
 \centering
\includegraphics[width=0.8\columnwidth]{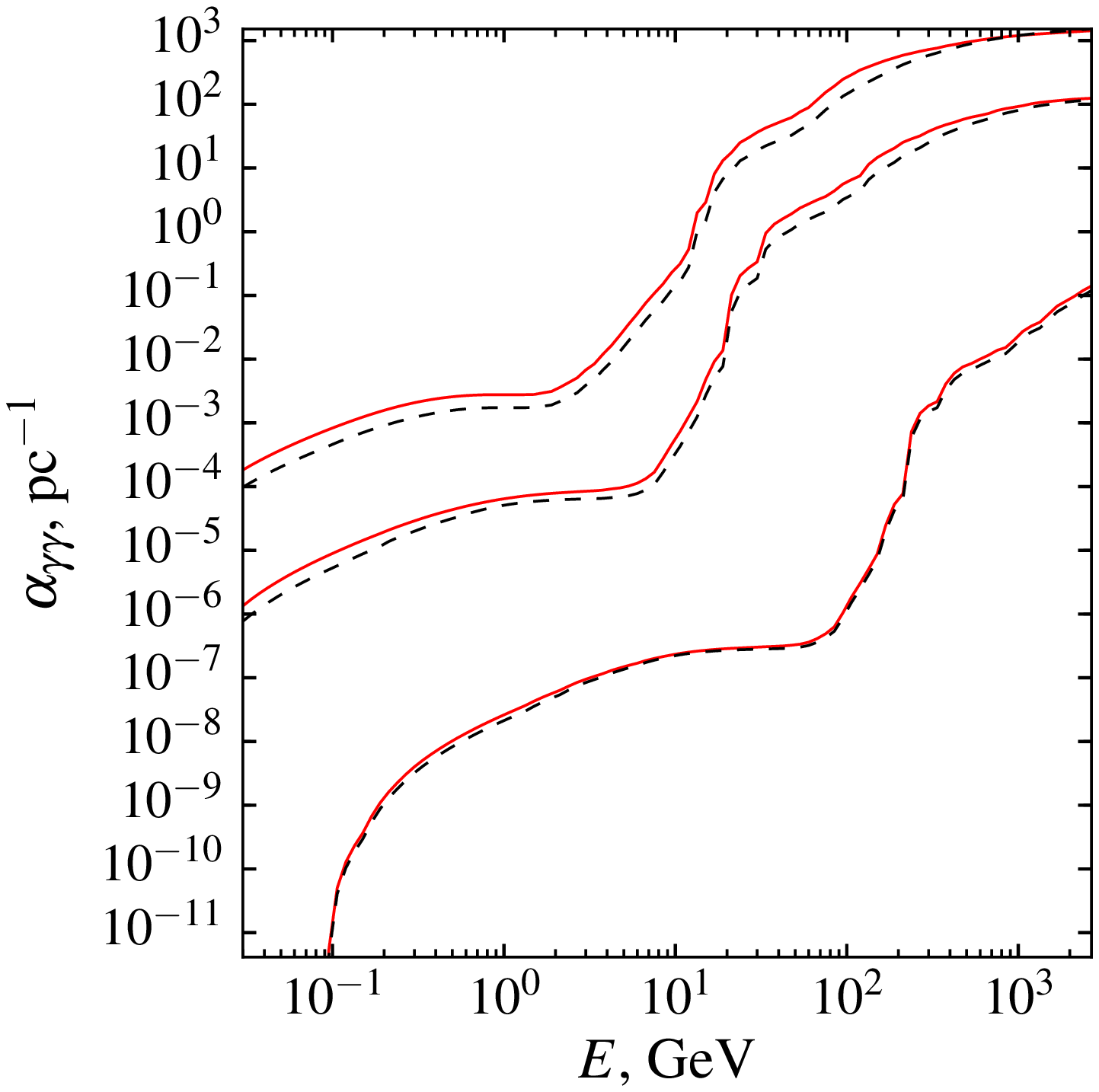}
\caption{Effects of the anisotropy upon the opacity coefficients shown for three distances of the gamma-ray source 
$H=0.01, 0.05$ and 0.3 pc (from top to bottom). 
The black dashed curves show the opacity created by isotropized BLR radiation, solid red curves
  correspond to standard anisotropic setup used in this paper, where we assume 
  disc-like geometry ($\theta_1=72^\circ$), $N_{\rm H} = 10^{23}\cmsq$, and 
  $n_{\rm H} =10^{11}\cmc$. 
}
\label{fig:scra}
\end{figure}

Anisotropy of the radiation emitted by BLR clouds is huge, but its effect upon the opacities is limited (see Fig.~\ref{fig:scra}). 
In general case, clouds seen at different angles contribute to the radiation received by the jet that diminishes the effect of anisotropy. 
For $H\gg R$, most of the clouds are either visible at angles close to $\pi/2$ (for disc-like geometry) or span a large range of inclinations. 
Anisotropy effects are the strongest for $H\ll R$ when only the inward-emitted flux contributes to the opacity. 
The intensity of the soft radiation and the optical depths in this case increase by about a factor of three with respect to the isotropic case. 

\begin{figure}
 \centering
\includegraphics[width=0.8\columnwidth]{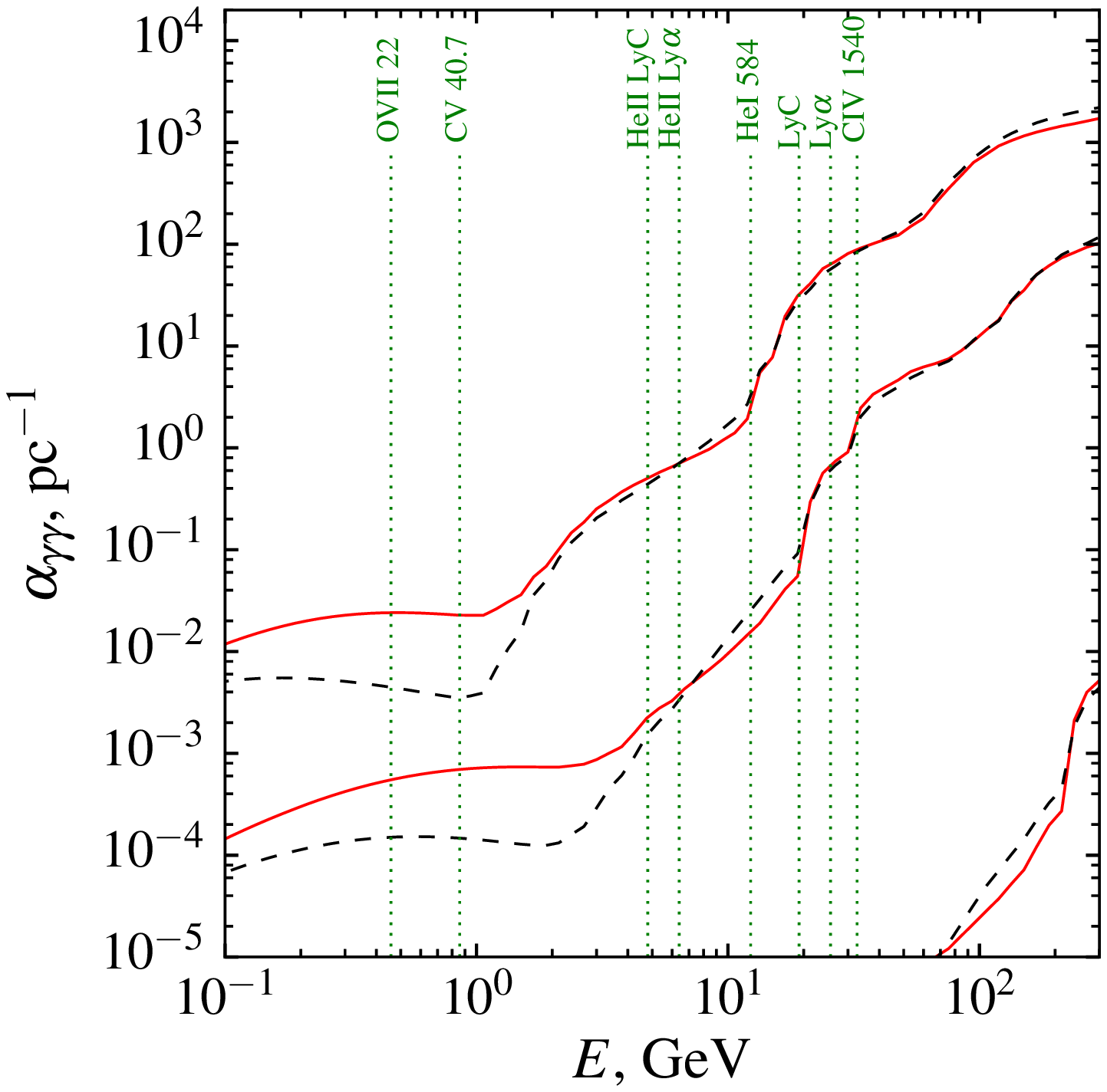}
\caption{Effects of  metallicity on the opacity coefficients for three distances  $H=0.01, 0.05$ and 0.3 pc (from top to bottom). 
The red solid and black dashed curves correspond to solar  and 10 times solar metallicities. 
We assume here $N_{\rm H} =10^{24}\cmsq$,  $n_{\rm H}=10^{10}\cmc$ and the disc-like geometry ($\theta_1=72^\circ$). 
}
\label{fig:mesmaps}
\end{figure}

\subsection{Effect of metallicity}

BLR are known to have on average super-solar metallicities \citep[see][]{Hammet,DWE,nitra}. 
Some objects are over-abundant in heavy elements by a factor of several, up to $Z\simeq 15Z_\odot$. 
It is unclear if this is a natural outcome of the metallicity gradient present in the host galaxies or a consequence of local intense star formation. 
For the hotter regions where the main coolants are carbon lines, higher metallicity does not strongly increase the intensities of metal lines, but the local temperature decreases thus increasing the intensity of Ly$\alpha$ that remains one of the main spectral details. 
However, Lyman emissions and ionization edges belonging to hydrogen- and helium-like carbon, nitrogen and oxygen become important in the low-energy part of the absorption spectrum. 
For a disc-like geometry, the absorption break moves to several GeV and can thus mimic intense Lyman \ion{He}{II} emission. 
In Fig.~\ref{fig:mesmaps}, we compare the opacities created by solar-metallicity gas and by a $Z=10Z_\odot$ gas. 
High metallicity changes the contribution of different CNO-element emissions and thus the absorption at $\sim 1\GeV$. 
The significantly lower low-energy absorption is due to lower temperature and ionization fraction of the higher-metallicity gas. 
As \cloudy\ does not take into account Thomson scattering on bound electrons, the real optical depth is thus underestimated in this region.

\section{Discussion}\label{sec:disc}

\subsection{The overall optical depth}\label{sec:disc:taumax}

The maximal possible optical depth to photon-photon interaction is 
\begin{equation}\label{E:taumax}
\tau_{\max}  \simeq
0.2\sigma_{\rm T} Q_{\rm ion} C /{4\pi cR} \simeq 0.04 C_{-1}\,L_{45} R_{\pc}^{-1}.
\end{equation}
If the radius $R$ is about the size of the BLR given by equation (\ref{E:Kaspi}), $\tau_{\max}  \simeq 4\sqrt{L_{45}}$. 
Luminosity of the central source can reach $L_{45} \gtrsim 300$ \citep{zhang15} that implies an optical depth up to $\sim 10^2$. 
Optical depths can also exceed this value if most of the radiation is produced in a more compact region. 

Inside the inner radius of the BLR, radiation field is practically uniform, and the optical depth saturates. 
At higher distances, radiation density decreases rapidly, and for $H\gtrsim R_{\max} $, the scaling being $\propto H^{-2}$. 
The decreasing range of incidence angles also affects the optical depth. 
If in equation~(\ref{E:tau}) we assume the intensity constant within the range of $-1<\mu<\mu_{\max} $, the {absorption coefficient becomes proportional to $\left(1+\mu_{\max} \right)^{3+\Gamma}$, where we approximate the $EI_E$ spectrum by a power-law $E^{-\Gamma}$. 
Because  the optical depth is given by the integral over height $\tau_{\gamma\gamma} \propto \int  \alpha_{\gamma\gamma}dh$ and 
because for large distances $1+\mu_{\max}\propto h^{-2}$, we get 
}
\begin{equation}\label{E:tauh}
\tau_{\gamma\gamma} \propto \int_H^{\infty} dh \left(1+\mu_{\max} \right)^{3+\Gamma}  \propto H^{-5-2\Gamma} .
\end{equation}
The maximum of the optical depth shifts to larger energies. 
At energies where the BLR spectrum has sharp features (e.g. Ly$\alpha$ line), the effective slope $\Gamma$ is very large. 
This leads to a very sharp dependence of the total optical depth on the distance $H$ in the $1\div 100\GeV$ range, 
where a number of strong lines contribute to the absorption.
 
This result also implies that the BLR radiation is only important for sources located at the distances $H\lesssim R_{\max} $ and the emission from the innermost parts of the BLR ($R\ll H$) is practically irrelevant for the gamma-ray absorption. 
The most crucial condition for efficient absorption is existence of photons with large incidence angles ($\mu \gtrsim 0$). 
If most of the broad line luminosity is emitted below the gamma-ray source, the absorption will be dominated by the outermost BLR clouds with an overall luminosity up to several orders of magnitude smaller than the total luminosity of the BLR. 
Therefore, the  UV spectrum required to explain the GeV-range absorption may indeed differ profoundly from the observed spectrum of the BLR. 
In fact, the absorption may be created by a distinct population of BLR clouds located along the jet axis and producing negligible total luminosity. 

\subsection{Breaks in gamma-ray spectra}\label{sec:disc:breaks}

Spectral data in the GeV range \citep{fermi10} show that the spectral shapes of blazar spectra deviate strongly from any simple power-law model. 
One of the reasons for this, but not necessarily the only one, is absorption to pair production.  
Gamma-ray absorption exists atop of an intrinsically curved spectrum of a roughly log-normal shape \citep{SP11}, that complicates the search for any unambiguous signature of absorption details. 

Blazars are variable sources and show strong spectral variability in the gamma-ray range \citep{fermi10,fermilarsson,SP11,foschini13,Willy14}. 
On one hand, it complicates the interpretation of the time-averaged spectra. 
On the other hand, studying spectral variability allows to set additional constraints upon the gamma-ray source and the putative absorption. 
One of the brightest and best studied objects is 3C\,454.3, where absorption details were reported in different brightness states \citep{ackermann, SP11}. Its spectrum seems to retain log-normal shape with a practically constant width but variable peak energy correlated with the flux. 
A couple of absorption details are identified with the edges produced by BLR emission lines, namely hydrogen and \ion{He}{II} Ly$\alpha$ and LyC, at about $5$ and $20\GeV$ in the blazar frame \citep{SP14}. 
Such breaks are also visible in the stacked redshift-corrected FSRQ spectra, and possibly in another object, 4C$+$21.35. 
{ Though there is a probable contamination of the spectra with the features
  of the {\it Fermi}/LAT response function, the spectral breaks are absent in
  the sky background, and therefore probably
  real. They are also probably connected to the  photon-photon absorption
  effects as in the spectra
  of BL~Lacs, the breaks are quite rare. The BL~Lac objects like
  AO\,0235$+$164 and PKS\,0426$-$380 \citep[see][]{fermi10,tanaka13} showing
  strong spectral breaks are low-synchrotron-peaked objects similar in their
  properties to FSRQ. In fact, the broad line luminosities of these objects
  are comparable to those of FSRQ. In particular, \citet{sbarufatti05}
  estimate the \ion{C}{III]}$\lambda$1909 and \ion{Mg}{II}$\lambda$2800
luminosities of PKS\,0426$-$38 as
about 5$\times 10^{42}$\ergl\ and 7$\times 10^{42}$\ergl, correspondingly,
similarly to the model BLR we considered in this paper and several times fainter
than for a bright FSRQ like 3C\,454.3 \citep[see for example][]{isler13}. Their
classification as BL~Lac is based upon the equivalent widths of the broad
emission lines (usually $EW\lesssim 5$\AAA)
and thus may result from the larger non-thermal flux, that is sensitive to orientation of the jet.
 }

As we have seen in Section~\ref{sec:res}, positions of the spectral breaks depend upon the geometry and may be parameterized by a single quantity $\mu_{\max}$ that depends upon the relative distance along the jet $H/R_{\max}$ and the half-opening angle of the BLR. 
The latter may be possibly constrained from the observations of the broad lines in AGN. 
Broad lines are best studied in relatively nearby Seyfert galaxies like NGC~5548 where it was possible to estimate the overall oblateness \citep{KZ13}. 
The shapes determined  using different lines differ significantly, from height-to-radius ratio of 0.2 for H$\beta$ to  the values exceeding unity for high-ionization lines. 
The real BLR seem to be not only non-spherical but stratified in meridional direction. 

\subsection{Very high energy emission}

The issue that requires special attention is existence of FSRQ clearly detectable at very high energies (VHE), at $E\gtrsim 100\GeV$. 
There are quite few objects of that kind, including PKS 1222$+$21 \citep{tanaka11,pks12}, 3C\,279 \citep{3c279} and PKS 1510$-$89 \citep{pks15}. 
In all these objects, the VHE spectrum conforms well with extrapolation of the GeV-range spectral energy distribution, if absorption by extragalactic background is corrected for. 
No additional absorption seems to be needed, meaning that the maximal internal
optical depth cannot be very high, $\tau_{\max} \lesssim 1$. However, for PKS 1222$+$21,
\citet{tanaka11} mentions spectral breaks at 1$\div$3\GeV\ observed during a strong gamma-ray
outburst of this object.

{ The optical emission lines of PKS 1222$+$21 allow to estimate the BLR
luminosity of this object as $L_{\rm BLR}\sim  5\times 10^{44}\ergl$
\citep{tanaka11}. Assuming that the thermal luminosity is ten times higher, 
we can estimate the size of the BLR as $0.02\div 0.03\pc$. The maximal optical
depth is then about 10 if the source is located inside the BLR and drops
rapidly with distance (see equation~(\ref{E:tauh})). 
For the shape of the ionizing spectrum we use in this work, the optical depth
changes by several orders of magnitude throughout the GeV range and reaches
its maximum around $0.1\div 1\TeV$.
If the optical depth is around unity in the TeV range, it should be several
orders of magnitude smaller in the GeV range that implies we should not be
able to see the absorption edges for this particular object, unless several different zones are
responsible for different parts of the gamma-ray spectrum. 

}
 
\section{Conclusions}

In this work we have considered the optical depths to pair production that a realistic BLR should produce. 
Certain simplifications were used such as uniform density within individual BLR clouds and uniform distribution of the clouds themselves within the volume they occupy. 
We took into account the geometry effects introducing an opening angle of the BLR distribution in space, and also anisotropic emission pattern of individual clouds, that should emit mostly backwards in the most prominent emission lines. 
The structure, physical conditions and the outcoming spectra of the BLR clouds were calculated using the photoionization code \cloudy.  

As many BLR show signatures of super-solar metallicity, we also calculated one
model with a metallicity ten times solar. High metallicity increases the
intensity of the soft X-ray metal lines, especially
\ion{C}{V}$\lambda$40.7. The opacity growth with energy in the
1$\div$100\GeV\ range becomes generally steeper. 

We find that the main effect of geometry is upon the positions of spectral
details in the absorption spectrum. 
The key quantity is the maximal incidence angle cosine $\mu_{\max} $. 
Moving the spectral breaks to higher energies makes the opacity in the
relevant GeV range much smaller, sometimes by several orders of magnitude, if
the geometry is changed from a complete sphere to a thin disc. 
At the same time, the optical depth at harder energies $E\gtrsim 100\GeV$
remains practically independent of energy and close to the maximal possible. 

{ Strong dependence on the incidence angle makes the gamma-ray opacity
  extremely sensitive to the position of the gamma-ray source as well as of
  the sources of soft photons. Most of the broad line radiation produced
  below the
  gamma-ray source (at $R\lesssim H$) remains invisible for the GeV-range
  photons. Therefore, the average BLR spectrum can not be used to predict the
  positions and strengths of the absorption edges. Instead, the absorption may
  be dominated by the radiation of a small part of the BLR located favourably
  with respect to the gamma-ray emission site. }

\section*{Acknowledgments} 

This work has been supported by the Academy of Finland grant 268740. JP
  also acknowledges support by the Foundations' Professor Pool and the
  Finnish Cultural Foundation.

\bibliographystyle{mn2ew}
\bibliography{mybib}

\appendix

\section{Lyman greenhouse effect}\label{sec:app:greenhouse}

\begin{figure*}
 \centering
\includegraphics[width=0.9\textwidth]{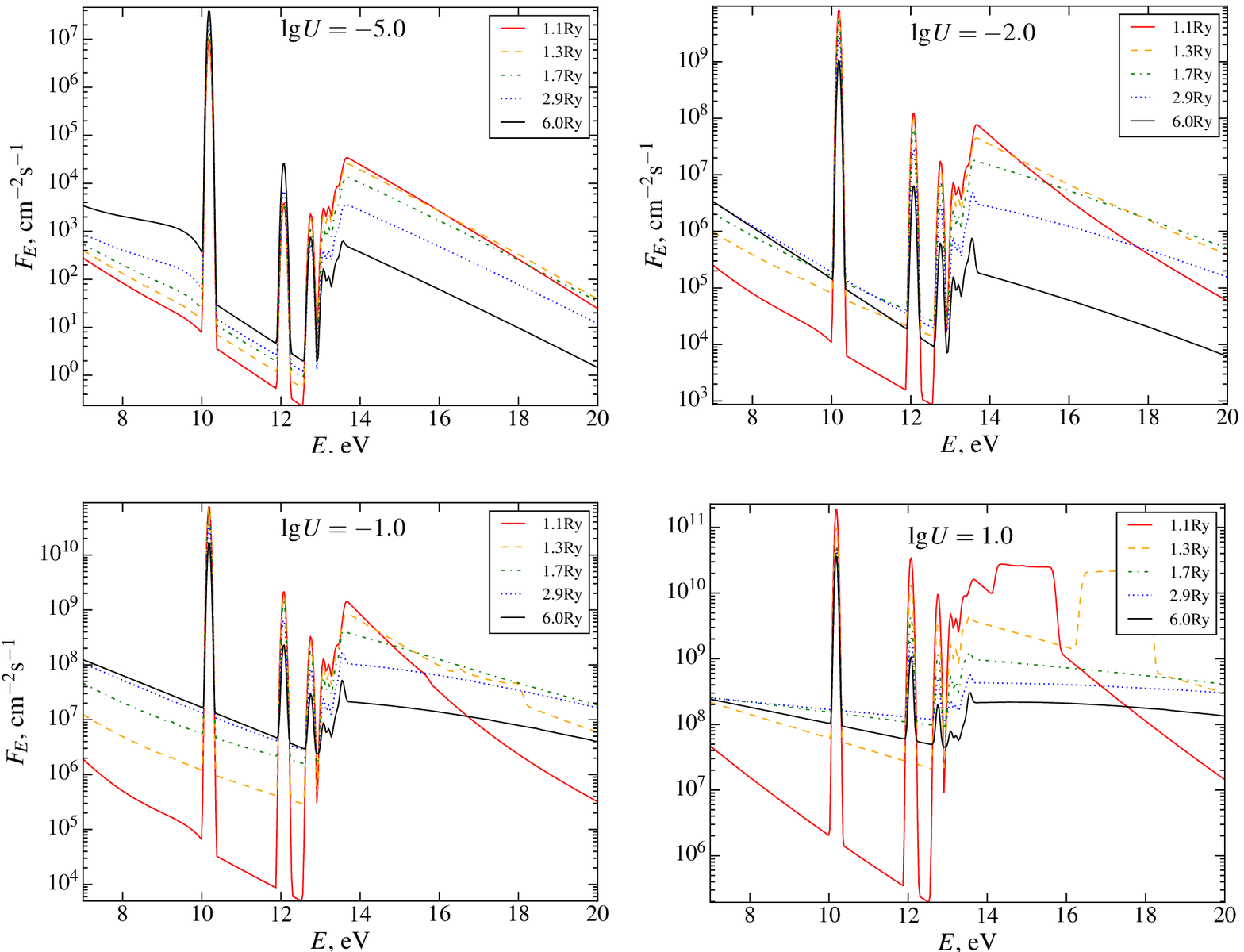}
\caption{ 
Reflected spectra of a monochromatic beam with different photon energies (given in the legend) for different ionization parameter values. 
}
\label{fig:mono}
\end{figure*}

\begin{figure*}
 \centering
\includegraphics[width=0.9\textwidth]{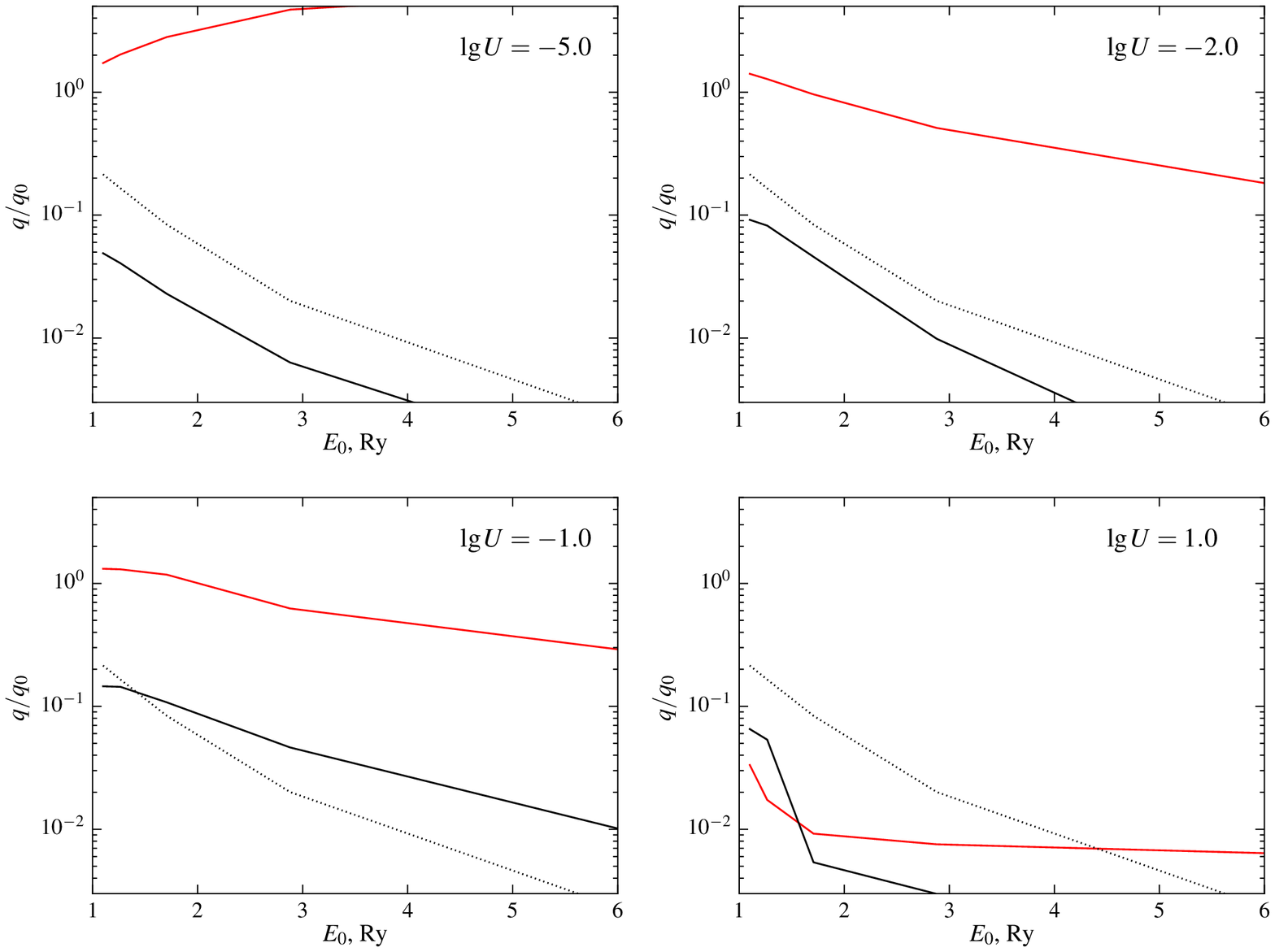}
\caption{ 
Ly$\alpha$ (red solid) and LyC (black solid; integrated between 1 and
1.5\Ry) photon fluxes in reflected spectrum as functions of beam energy. 
Photon flux density in the ionizing beam was held constant. 
Black dotted line is the incident LyC photon flux multiplied by $\dfrac{1}{2} \left( 1 + \left(\frac{\Ry}{\langle E\rangle}\right)^3\right)^{-1}$. 
}
\label{fig:qas}
\end{figure*}

As we have seen in Section~\ref{sec:blr}, Ly$\alpha$ dominates the reflected spectrum, being much stronger than, for instance, the Lyman recombination continuum. 
At least to some degree, it is an outcome of ionization by hard radiation. 

To illustrate this, let us consider monochromatic ionizing beam with photon energy $E_0$ and initial number flux density $q_0$. 
After travelling a distance of $l$ through a neutral medium with constant hydrogen density $n$, it is attenuated (if $E_0 \geq 1 \Ry$) exponentially as:
\begin{equation}
q(l) = q_0 e^{-\sigma_1\left(\frac{\Ry}{E_0}\right)^3 n l},
\end{equation}
where $\sigma_1\simeq 6.3\times 10^{-18}$~cm$^2$ is the photoionization cross-section near the ionization edge. 
The rate of primary photoionizations (number of ionizations per cubic centimeter per second) is thus
\begin{equation}\label{E:N}
\dot{N}(l)= \sigma n q(l) = \sigma_1\left(\frac{\Ry}{E_0}\right)^3 n q_0
e^{-\sigma_1\left(\frac{\Ry}{E_0}\right)^3 n l}.
\end{equation}
Ionization is balanced by recombinations upon the ground and excited levels. 
Recombinations on the ground level produce LyC quanta, subordinate recombinations produce optical and IR recombination continua and Lyman series lines. 
Probability of emitting a Lyman-continuum photon is about 0.5 \citep{OTS}.
All the Lyman-continuum quanta that reach the observer  are formed at small optical depths. 
The local spectrum of these secondary Lyman continuum quanta is thermal, and the mean energy $\langle E\rangle \sim 1\Ry$ that is for most cases much smaller than $E_0$. 
The number flux of secondary LyC quanta escaping the cloud is
\begin{eqnarray}\label{E:qsec}
\displaystyle q_{\rm C}& =& \int_0^{+\infty}e^{-\sigma_1\left(\frac{\Ry}{\langle
    E\rangle}\right)^3 nl} n \sigma q(l) dl  \nonumber \\
\displaystyle  &=&  nq_0 \sigma_1 \left(\frac{\Ry}{E_0}\right)^3 \int_0^{+\infty}e^{-\sigma_1\left(\frac{\Ry}{\langle
      E\rangle}\right)^3 nl} e^{-\sigma_1\left(\frac{\Ry}{E_0}\right)^3 nl} dl \nonumber \\ 
\displaystyle  &=& \dfrac{q_0}{1 + \left(\frac{E_0}{\langle
    E\rangle}\right)^3}.   
\end{eqnarray}

If the ionizing continuum is hard enough ($E_0 \gg \langle E\rangle \sim \Ry$), the number of LyC quanta in reflected spectrum decreases as $\propto E_0^{-3}$, while the number of Ly$\alpha$ should remain more or less the same if collisional de-excitation and two-photon emission do not play important role. 
Also, destroyed LyC quanta are converted to higher-order recombination continua plus Lyman series lines, mostly Ly$\alpha$, that increases the contrast between the line and the recombination continuum even more. 

To check this more rigorously, we calculated a simple \cloudy\ model of a thick plane-parallel pure hydrogen slab irradiated by an effectively monochromatic beam with photon energy $E_0$. 
We used the {\tt laser} command to reproduce a narrow-spectrum radiation source with an energy spanning the range from 1.1 to 6\Ry. 
The ionization parameter was varied independently between $10^{-5}$ and 1. 

In Fig.~\ref{fig:mono}, we show the reflected emission line spectra for different beam energies. 
Integrating the relevant parts of the spectrum, one can estimate the number of LyC and Ly$\alpha$ photons. 
In Fig.~\ref{fig:qas}, the number of Ly$\alpha$ quanta is essentially the same as the number of LyC quanta in the ionizing continuum, especially for higher ionization parameters. 
For lower ionization parameters, $U\lesssim 0.01$, Ly$\alpha$ is the primary coolant, and the number of quanta emitted in the line scales with the beam energy. 
For high ionization parameters, the gas becomes hot and nearly totally ionized, and the reflected spectrum is dominated by electron scattering. 
For $U=10$, in particular, the contribution of the incident continuum is clearly seen in the reflected spectrum (Fig.~\ref{fig:mono}, lower right panel).

\section{The details of the {\tt Cloudy} model}\label{sec:app:cloudy}

This is a sample \cloudy~13.03 input used to calculate the nebular spectra of BLR clouds:

{\tt 
\noindent
interpolate (0.000093 -5.0) (0.0093 0) (4.0 -2.636)\\
continue (30.0 -4.3837) (7350. -6.7729) (73500. -7.773)\\
luminosity 45.0 total\\
radius 0.138949549437 parsec linear\\
hden 11.0 log\\
abundances HII region no grains\\
atom FeII levels 100\\
covering factor 1.\\
stop column density 23.0\\
iterate to convergence\\
save continuum ".con" last iteration units= eV\\
save line optical depths, limit=0 ".tau"\\
save cooling ".cool" last iteration\\
save temperature ".temp" last iteration\\
}

The {\tt interpolate..continue } command interpolates through several points with power laws and produces the ionizing continuum shape introduced by equation~(\ref{E:isp}). 
The energies of the turning points are given in Rydbergs. 
The lower break here is at 100$\mu$. 
The luminosity is set to $10^{45}\ergl$, the radial distance is given in parsecs, and hydrogen density is defined by the {\tt hden} parameter as a logarithm. 
Abundance set {\tt HII region} is one of the standard solar-metallicity sets used by \cloudy~\citep{cloudy13}. 
However, considering the effect of high metallicity, we used the {\tt abundances starburst, Z=10} instead that uses the predictions of starburst chemical evolution simulations by \citet{HF93}. 
Abundances of different elements scale non-linearly and should be estimated using comprehensive modelling of star formation and recycling of chemically enriched matter. 
However, the chemical composition effects are small enough to neglect the differences between individual abundance sets at $Z\simeq Z_\odot$.
Geometry is open, and the {\tt covering factor 1} command only ensures that the output flux is not rescaled by any kind of multiplier containing the covering factor.
The structure of the nebula is calculated iteratively until convergence. 
The last several lines define different kinds of output including  continuum, optical depths, cooling and temperature structure of the model slab.

\section{\cloudy\ versus \xstar}\label{sec:app:xstar}

\begin{figure}
 \centering
\includegraphics[width=0.8\columnwidth]{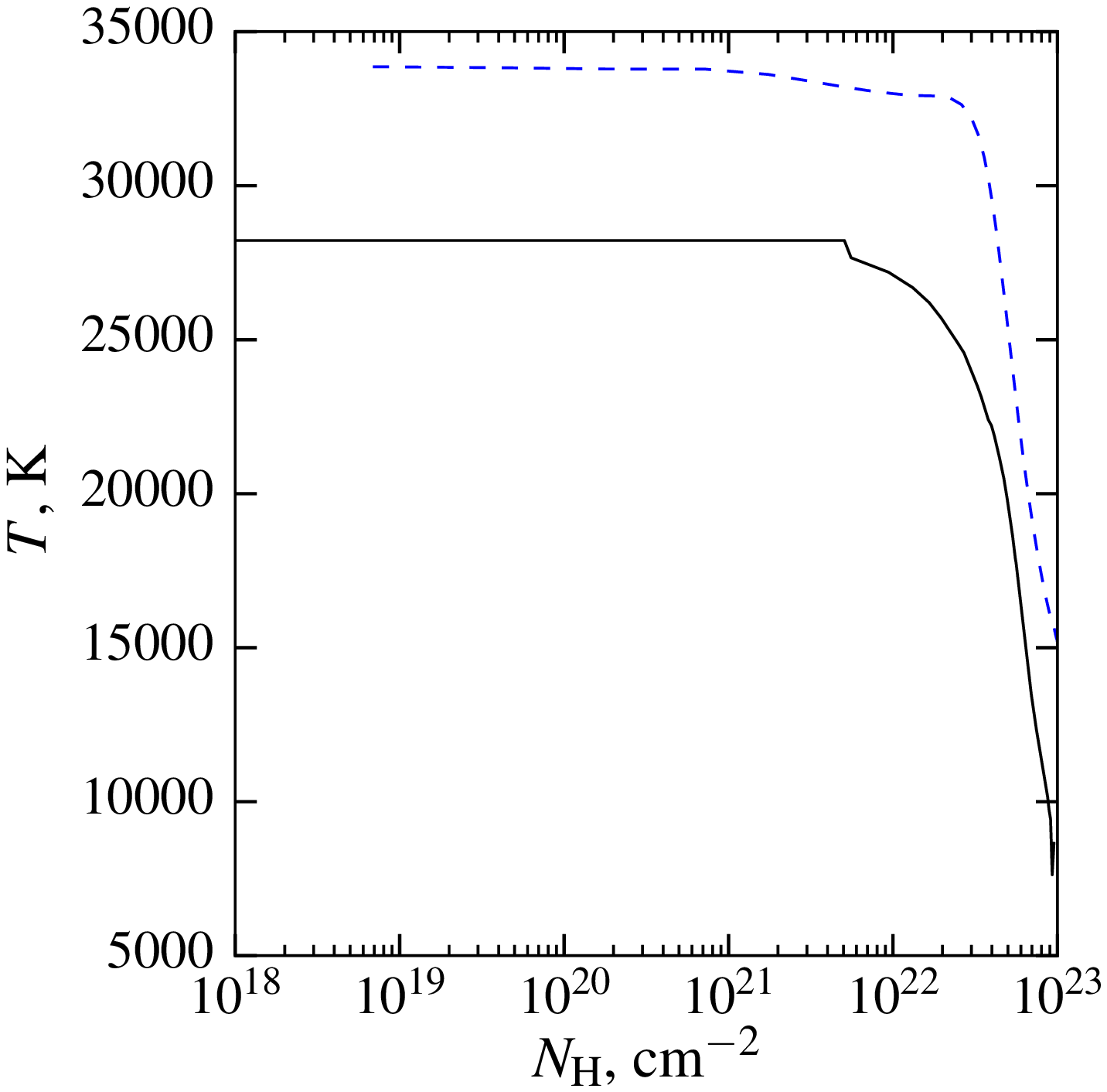}
\caption{ 
Temperature slices of the \xstar\ (black solid) and \cloudy\ (blue dashed) models of  pure hydrogen slab illuminated by a power-law continuum. 
Here density is $10^9\cmc$ and ionization parameter $U\simeq 0.19$. 
}
\label{fig:XC:temp}
\end{figure}

\begin{figure}
 \centering
\includegraphics[width=0.8\columnwidth]{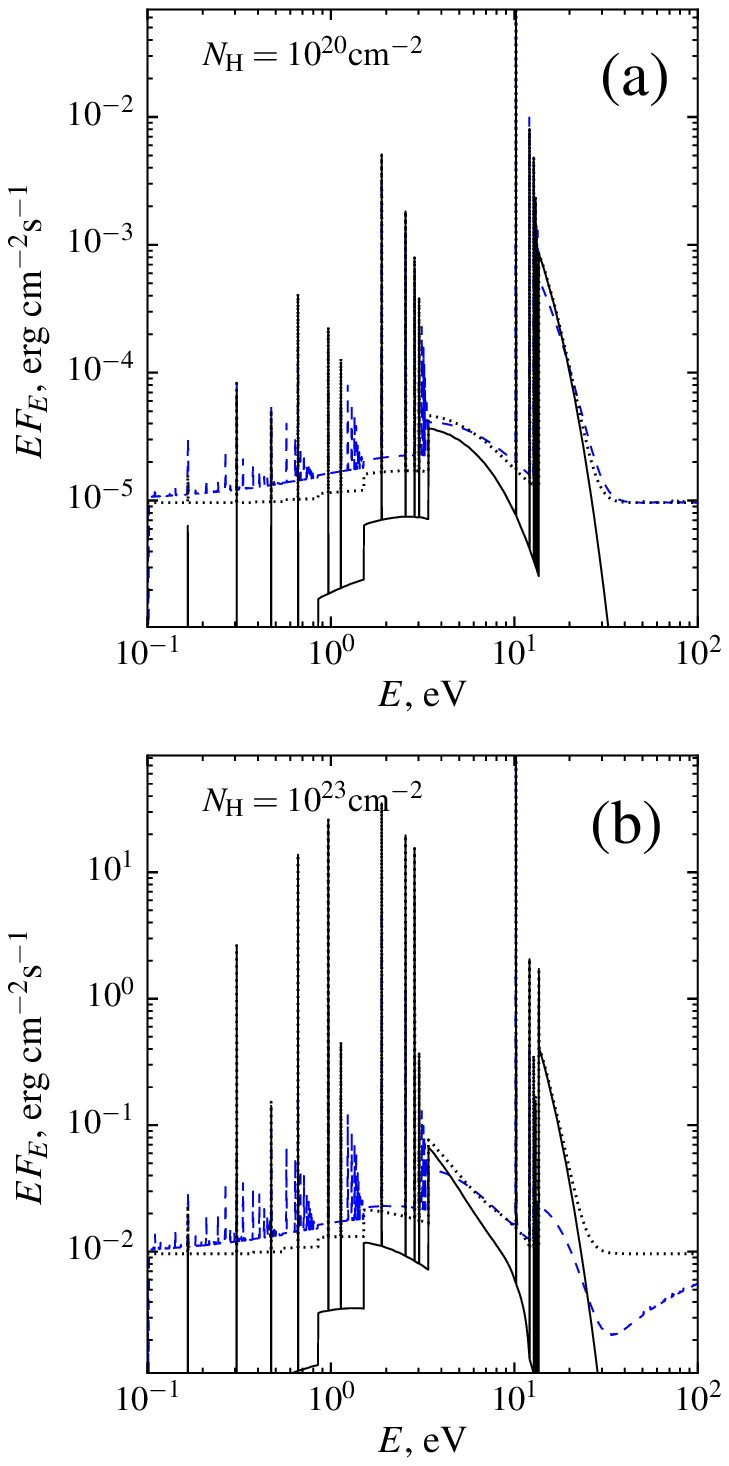}
\caption{ 
Reflected spectra produced by \cloudy\ and \xstar\ pure-hydrogen models for different column densities:  (a) ``nebular'' column density $10^{20}\cmsq$  and  (b) typical BLR cloud column density $10^{23}\cmsq$ . 
The blue dashed line is the \cloudy\ output and the raw \xstar\ spectra are shown with the black solid line. 
The black dotted curves show \xstar\ spectra plus an incident continuum multiplied by the Thomson optical depth ($\tau_{\rm T} \simeq 0.0665 N_{\rm H,23}$). 
}
\label{fig:XC:spectra}
\end{figure}

Though different photoionization codes proved to provide consistent results for the parameter ranges of ordinary \ion{H}{ii}-regions \citep{pequignot}, including planetary nebulae and even, to some extent, the narrow-line regions of active galaxies,  to our knowledge they have not been  compared  in the extreme regime of BLR conditions. 
The challenges of BLR modelling include not only high densities but also high optical depths in resonance lines and recombination continua. 
Optical depths for Ly$\alpha$ and LyC quanta reach about $10^6$ for the hydrogen column densities typical for BLR clouds. 

We have calculated a series of \xstar\ and \cloudy\ models with identical ionizing spectra and physical conditions. 
A power-law flat spectrum ($E L_E = \mbox{const}$) between $1$ and $10^3\Ry$ was adopted as one of the basic models used by \xstar, most relevant for modelling AGN emission. 
To exclude the effects of abundances and atomic data, and also to speed up the calculation, pure hydrogen models were calculated.

Heating/cooling balance is solved in a similar way. In real BLR, the main coolants are multiply-scattered Ly$\alpha$ and permitted metal lines including CIV$\lambda$1549 and MgII$\lambda$2800, but pure-hydrogen models are cooled mainly by Ly$\alpha$ and recombination continua. 
The resulting temperatures are around $3\times 10^4\rm K$ according to both codes (see Fig.~\ref{fig:XC:temp}), but \cloudy\ results are systematically higher, probably because of lower escape probability of recombination photons (see below). 

\xstar\ does not include the radiation scattered by free electrons in the diffuse continuum emission. 
Adding the incident continuum multiplied by the estimated optical depth to electron scattering makes the outputs of the two codes in the X-ray and infrared ranges much closer to each other (see Fig.~\ref{fig:XC:spectra}). 
In fact, both codes do not include the X-ray radiation scattered by bound electrons. 
However, \cloudy\ takes ``bound Compton'' into account as a source of heating and absorption.\footnote{See discussion
in the \cloudy\ discussion group,
\url{https://groups.yahoo.com/neo/groups/cloudy_simulations/} \url{conversations/topics/2729}} 
The lack of scattered X-ray emission is visible, for instance, in the lower panel of Fig.~\ref{fig:ltot}. 

UV and optical radiation are also well consistent with each other for smaller column densities. 
However, at large column densities, code predictions differ a lot. 
In particular, \xstar\ always predicts strong Lyman continuum emission, and in general stronger recombination continua that \cloudy. 
In Fig.~\ref{fig:XC:spectra}, we show an example of reflected spectra for two pure-hydrogen spectral models with different column densities, $N_{\rm H} = 10^{20}\cmsq$ (a) and $N_{\rm H} =10^{23}\cmsq$ (b).   
Density was set to $10^9\cmc$, and the radiation source with a luminosity between 1 and  $10^3\Ry$ equal to $10^{45}\ergl$ placed at the distance of 0.1\pc, that  implies an ionization parameter $U\simeq 0.19$. 
The optical depths near the Lyman edge are about $600$  and  $6\times 10^5$ for the thinner  and the  thicker model, respectively. 
Both values are much larger than unity, but the latter lies in the range typical for ordinary \ion{H}{ii}-regions and thus well studied both
observationally and numerically with different photoionization codes. 
While the shape and the intensity of the Lyman continuum are similar for the small column density, for  the higher column density value typical for BLR, the difference is larger than an order of magnitude. 
The most probable origin of this discrepancy is in the approximation used for escape probability of Lyman-continuum photons (formula (5) in \citealt{xstar}) that was checked only for the environment of ordinary \ion{H}{ii}-regions having Lyman continuum optical depths of hundreds to thousands. 
For the optical depths about a million, more realistic radiative transfer treatment should be used. 
In \cloudy, either a much more accurate ``modified on-the-spot'' approximation is used, or the ``outward-only'' approximation \citep{oonly}. 
Both provide results consistent within several per cent. 
As we show in Appendix~\ref{sec:app:greenhouse}, there are physical reasons for the reflected continuum to be much fainter than Ly$\alpha$, and \cloudy\ results, unlike the results produced by \xstar, appear to be consistent with this expectation.

\label{lastpage}

\end{document}